\documentclass[onecolumn]{emulateapj}
\usepackage{graphicx}
\usepackage{color}
\usepackage{multirow}
\usepackage{setspace}

\usepackage{epsfig}
\usepackage{natbib}
\usepackage{wasysym}
\usepackage{verbatim}

\newcommand{\PS}{power spectrum}

\newcommand{\PA}{power spectra}

\newcommand{\PL}{power-law}
\newcommand{\PLs}{power-laws}

\newcommand{\Fps}{Fourier \PS}
\newcommand{\mFps}{mean Fourier \PS}

\newcommand{\Fpa}{Fourier \PA}
\newcommand{\mFpa}{mean Fourier \PA}

\newcommand{\BF}{ }

\shorttitle{Coronal Fourier power spectra}
\shortauthors{Ireland et al.}

\begin{document}


\title{Coronal Fourier power spectra: implications for coronal
  seismology and coronal heating}


\author{J. Ireland}
\affil{ADNET Systems, Inc.,NASA Goddard Space Flight Center, MC. 671.1, Greenbelt, MD 20771, USA.}

\author{R. T. J. McAteer}
\affil{Department of Astronomy, New Mexico State University, Las
  Cruces, NM.}

\author{A. R. Inglis}
\affil{NASA Goddard Space Flight Center, Greenbelt, MC. 671.1, MD, 20771, USA.}
\affil{Physics Department, The Catholic University of America, Washington, DC, 20664, USA.}


\begin{abstract}

  {\BF The dynamics of regions of the solar corona are investigated
    using {\BF Atmospheric Imaging Assembly (AIA)} 171\AA\ and
    193\AA\ data.  The coronal emission from the quiet Sun, coronal
    loop footprints, coronal moss, and from above a sunspot is
    studied.  It is shown that the \mFpa\ in these regions can be
    described by a \PL\ at lower frequencies that tails to flat
    spectrum at higher frequencies, plus a Gaussian-shaped
    contribution that varies depending on the region studied. This
    Fourier spectral shape is in contrast to the commonly-held
    assumption that coronal time-series are well described by the sum
    of a long time-scale background trend plus Gaussian-distributed
    noise, with some specific locations also showing an oscillatory
    signal.  The implications of this discovery to the field of
    coronal seismology and the automated detections of oscillations
    are discussed.  The \PL\ contribution to the shape of the Fourier
    power spectrum is interpreted as being due to the summation of a
    distribution of exponentially decaying emission events along the
    line of sight.}  This is consistent with the idea that the solar
  atmosphere is heated everywhere by small energy deposition events.
\end{abstract}





\section{Introduction}\label{sec:int}
Coronal seismology is the study of oscillatory phenomena in the solar
corona.  First suggested by \cite{1970PASJ...22..341U}, the practice
of coronal seismology attempts to locate oscillatory phenomena in the
solar corona, identify the wave modes present, and then use
theoretical descriptions of those wave modes to infer the physical
conditions of the solar corona.  The practical application of coronal
seismological ideas began in earnest with observations of wave modes
{\BF described in \cite{1998ApJ...501L.217D} and
  \cite{1999SoPh..186..207B} from data captured by Extreme Ultraviolet
  Telescope (EIT, \citealp{1995SoPh..162..291D}) on board the Solar and Heliospheric
  Observatory (SOHO, \citealp{1995SoPh..162....1D}), and also by
  \citep{1999ApJ...520..880A} using data from the Transition Region
  and Coronal Explorer (TRACE, \citealp{1999SoPh..187..229H}).  Since these initial works,
  many papers have been published describing oscillatory phenomena in
  the corona and their interpretation in terms of theoretically known
  wave modes.  The review articles \cite{lrsp-2005-3} and
  \cite{2012RSPTA.370.3193D} refer to many more articles on both
  observational and theoretical coronal seismology}.

Oscillatory phenomena in the solar corona are relatively rare, both
spatially and temporally.  Two types of oscillatory motions have
received a lot of attention.  The first is known as a transverse
oscillation, since the loop motion is approximately transverse to the
extent of the loop \citep{1999Sci...285..862N}.  These motions appear
to be triggered by nearby flaring events, and are identified with the
fundamental harmonic of the kink mode of the flux tube that comprises
the coronal loop.  The second type is known as a longitudinal
oscillation, or propagating disturbance (PD).  These waves propagate
along flux tubes, and have been identified as slow mode
magnetohydrodynamic (MHD) magneto-acoustic waves (although alternative
interpretations do exist that do not invoke oscillatory phenomena, for
example, \citealp*{0004-637X-722-2-1013} and
\citealp*{2041-8205-727-2-L37}).  

Other types of oscillatory behavior have been observed.
\cite{2005ApJ...620.1101M} demonstrate the presence of oscillatory
power with periods in the range of 40-80 seconds using high-cadence
H$\alpha$ blue wing observations of a Geostationary Operational
Environmental Satellite (GOES) C9.6-class solar flare obtained at Big
Bear Solar Observatory using the Rapid Dual Imager
\citep{2000SoPh..193..259P, 2007A&A...473..943J}. The measured
properties of these oscillations are consistent with the existence of
flare-induced acoustic waves within the overlying loops.  In addition,
\cite{2013ApJ...772...54A} show that the oscillatory power in the
lower solar atmosphere of active regions increases in response to
flare events at distant solar locations.

The relative scarcity of observations of coronal oscillations has
prompted efforts to create automated oscillation detection algorithms.
The need for an automated oscillation detection algorithm has been
exacerbated by the large and rapidly growing amount of {\BF Solar
  Dynamics Observatory (SDO, \citealp{2012SoPh..275....3P})
  Atmospheric Imaging Assembly (AIA, \citealp{2012SoPh..275...17L})}
data.  There have been several attempts at designing such algorithms.
\cite{2004SoPh..223....1D} describe the design of an automated
oscillation-detection algorithm based on wavelet analysis. The
algorithm finds significant wave packets ranging from single to
multiple wave cycles in duration, by a wavelet power/confidence level
comparison against the null hypothesis that a given time series is
Gaussian-distributed noise.  Pixels with significant oscillatory
content are grouped manually.  \cite{2007SoPh..241..397N} take a
similar approach, using a thresholded fast Fourier transform to find
locations in TRACE data that may support an oscillatory signal. The
threshold level is defined as three to four times the average fast
Fourier transform (FFT) power; if the maximum FFT power is above this
level then the frequency at which that power occurs is assumed to be
real. \cite{2010SoPh..264..403I} assume that an oscillatory signal is
present in each location in the datacube and then use a Bayesian-based
treatment to calculate the probability that the frequency has a
particular value.  This algorithm relies on subtracting a background
trend from the time-series being tested, and assumes that the noise is
Gaussian distributed.

The above algorithms precipitate detections based on the strength of
the signal of an oscillation on a pixel by pixel basis, and then group
these pixels together using some set of criteria.  The automated
detection algorithm of \cite{2008SoPh..248..395S} begins by
identifying candidate oscillatory regions as those whose time-series
of emission show a high variance \citep{2003SoPh..213..103G} (under
the implicit assumption that the time-series has an approximately
constant mean value).  Detections are based on wavelet filtering of
these regions in all three dimensions of the data cube simultaneously,
with the implicit assumption that the noise is Gaussian-distributed.
{\BF The algorithm of \cite{2008SoPh..252..321M} differs from the
  previous algorithms: it begins by Fourier transforming the entire
  data cube, and calculating the coherence of neighboring pixels in
  narrow frequency bands, and filtering the results. The null
  hypothesis here is that pairs of time series that have low coherence
  can be rejected as being part of an oscillatory group.  This method
  uses a symmetric filter to estimate the coherence, in that
  oscillatory power from either side of the central search frequency
  contributes the same weight to the final result.  There is an
  implicit assumption here that the power on either side of the
  central search frequency is approximately the same.  In this
  algorithm, the spatial extent of the coherence of the oscillatory
  signal is the key identifier of a wave process.}

All these detections algorithms, and many detections in the literature
make either the explicit or implicit assumption that the observed
time-series consists of a background trend of some kind and
(approximately) Gaussian-distributed noise which must be taken in to
account in order to test for the presence of an oscillatiory signal
{\BF \citep{2004SoPh..223....1D, 2007SoPh..241..397N,
    2008SoPh..248..395S, 2010SoPh..264..403I, 2013SoPh..286..405C}.}
{\BF However, two recent papers cast doubt on this assumption.
  \cite{2014AA...563A...8A} integrate emission over small portions of
  active regions and the quiet Sun as observed in the 195\AA\ passband
  images from EIT and show that the resulting time-series has an
  approximate \PL\ \Fps\ over the frequency range 0.01 - 1 mHz.
  \citet{gupta2014} showed \PL\ \PA\ in the intensity at six single
  points in AIA 171\AA\ coronal plumes extending over the frequency range
  $0.3\rightarrow 4.0$ mHz.}

{\BF The aim of this paper is to investigate the nature of the Fourier
  power spectrum in regions of interest to coronal seismology, to look
  for evidence that the above assumption is appropriate.}  The nature
of coronal emission has implications for coronal seismology, and may
be indicative of the nature of the fundamental energy deposition that
keeps the corona hot.  Section \ref{sec:obs} describes the
observations used, while Section \ref{sec:anal} describes the method
of analysis employed and presents the results.  Section
\ref{sec:discuss} discusses the implication of these results for
coronal seismology and the heating of the solar atmosphere.  Section
\ref{sec:conc} summarizes the discussion and outlines future work.

\section{Observations}\label{sec:obs}
{\BF AIA} observes images of the Sun in multiple wavebands
simultaneously and continuously, at high cadence (12 seconds) and at
high spatial resolution (0.6 arcseconds per pixel).  These data enable
detailed comparisons between phenomena observed in multiple wavebands,
and are therefore ideal for assessing the frequency content of the
solar atmosphere.  {\BF AIA} data was acquired using the Lockheed Martin
Solar Astrophysics cutout service at {\it
  http://lmsal.com/get\_aia\_data/}.  Six hours of image data in the
time range 2012-09-23 00:00 - 06:00 UT, at the full 12 second cadence,
in the 171\AA\ and 193\AA\ wavebands is used.  These wavebands were
selected because previous studies in similar wavebands (TRACE
171\AA\ and 193\AA) show the presence of coronal oscillations.  {\BF
  The selected area covers a single $\beta\gamma$ active region, NOAA
  AR 11575}.  This area was selected for two reasons.  First, it is
quiescent in that there were no X-ray flares or other large-scale
disturbances (for example, filament eruptions or rearrangments of long
bright coronal loops) over the spatial extent of the cutout within the
time range (Figure \ref{fig:loc171193}).  Therefore, any variations in
the plasma properties cannot be ascribed to large-scale disturbances.
Second, the selected cutout region contains a sunspot, and sunspots
are known to be locations over which three minute oscillations have
been detected by other instruments \citep{2002A&A...387L..13D}.

The data was prepared for analysis using the SolarSoft / IDL routines
READ\_SDO and AIA\_PREP.  These procedures convert the downloaded
level 1.0 FITS cutout files to level 1.5 FITS files.  This involves
translating, scaling and rotating the images so they have the same
sun-center, image scale with solar north in the same direction.  All
further processing from this point was performed in the SunPy 0.4
(http://www.sunpy.org) data analysis environment.  Images were
de-rotated to compensate for solar rotation, based on the center of
the field of view of each image.  After de-rotation, each image layer
was coaligned with a reference layer, defined to be half way through
the dataset, in this case at approximately 03:00 UT (Figure
\ref{fig:loc171193}).  This process is required since the solar
de-rotation function used does not fully compensate for all the solar
rotation. {\BF Coalignment was implemented using a template matching
  technique \citep{lewis1995fast} implemented by the {\it
    scikit-image} \citep{Vanderwalt2014} routine {\it
    match\_template}.  A rectangular template sub-image with sides one
  half the size (in pixels) of the reference layer is defined.  The
  coalignment algorithm finds the location of the best match of this
  template with each layer.  The image layer is then shifted (using
  sub-pixel shifts) according to the location of the best match of the
  template with the image.  This process removes residual bulk motions
  of the images.}

\section{Analysis}\label{sec:anal}
\begin{figure}
\centerline{
\plottwo{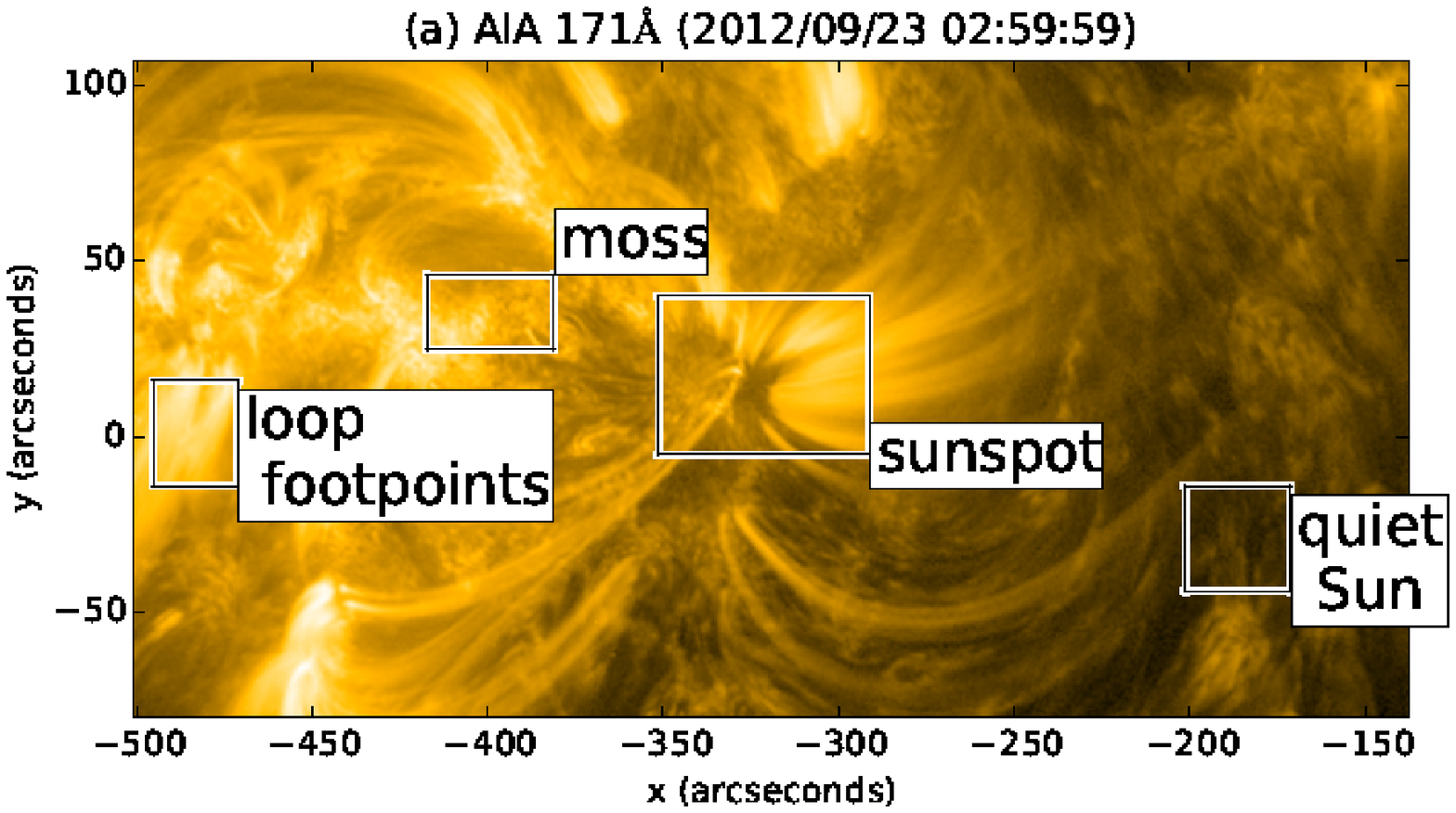}{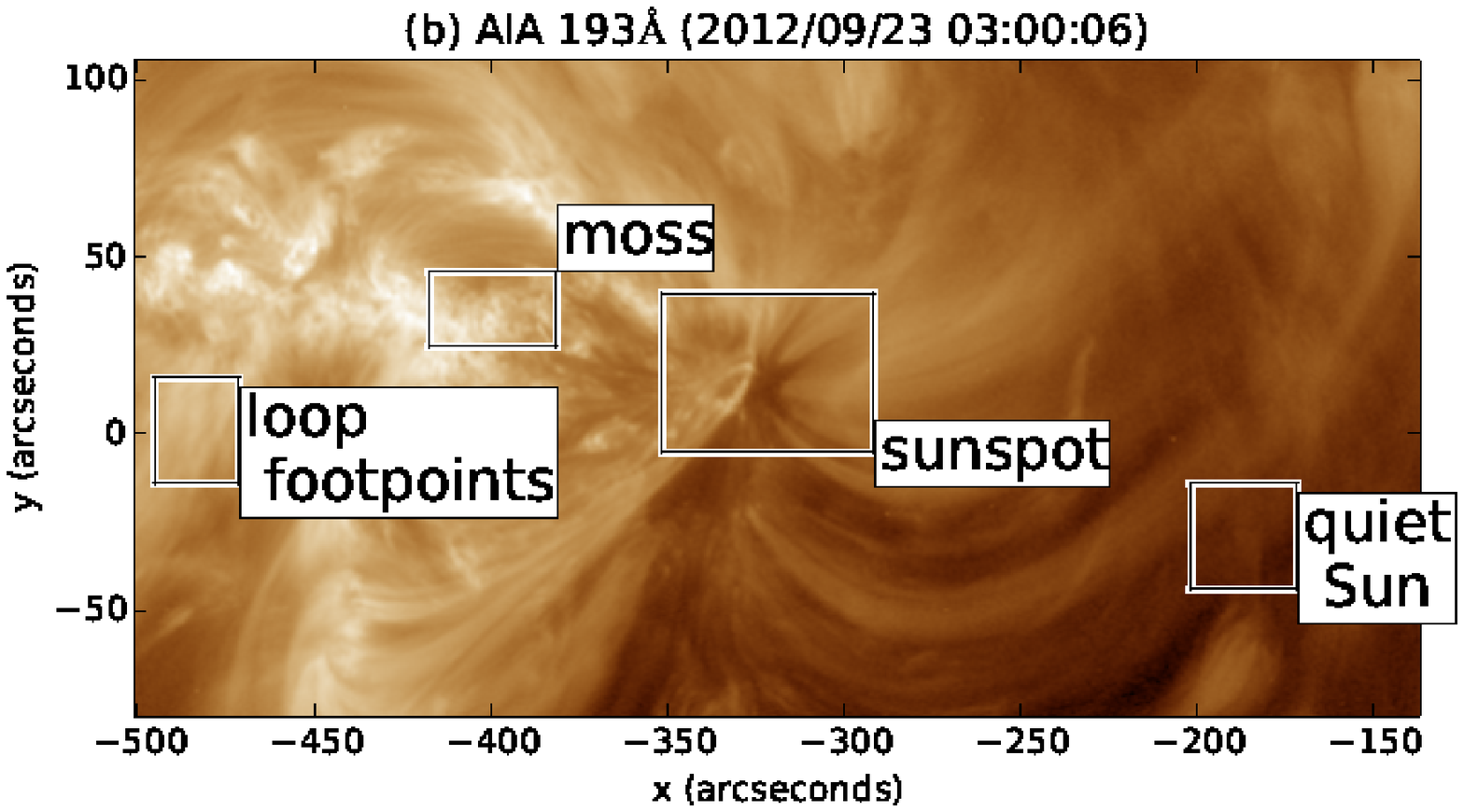}
}
\caption{Locations of the regions analyzed in each of AIA 171\AA\ and
  193\AA, overlaid on sample images taken halfway through the six hour
  span used to form the time-series analyzed.  The spatial extent of
  the data is (-501'', -138'') in the x-direction, and (-80'', 107'')
  in the y-direction, in heliocentric cartesian co-ordinates.  The
  images show NOAA AR 11575 and some surrounding area.}
\label{fig:loc171193}
\end{figure}

We wish to analyze the frequency content of these resulting datacubes
as a function of spatial location.  Four locations were chosen in the
data representing four different types of physical locations in the
solar atmosphere.  These were a quiet Sun region, a region that lies
on top of a sunspot, a footpoint region, and a coronal moss region.
Figure \ref{fig:loc171193} shows the location of the regions on sample
images from the time range considered. {\BF These locations, selected
  by visual inspection, were chosen because they are representative of
  commonly observed structures in the corona.  Further,
  oscillatory power has been identified previously over sunspots and
  loop footpoints \citep{2002A&A...387L..13D} and moss structures
  \citep{2003ApJ...595L..63D}.  A quiet Sun region is also included
  for comparison.}

\begin{figure}
\centerline{
\plottwo{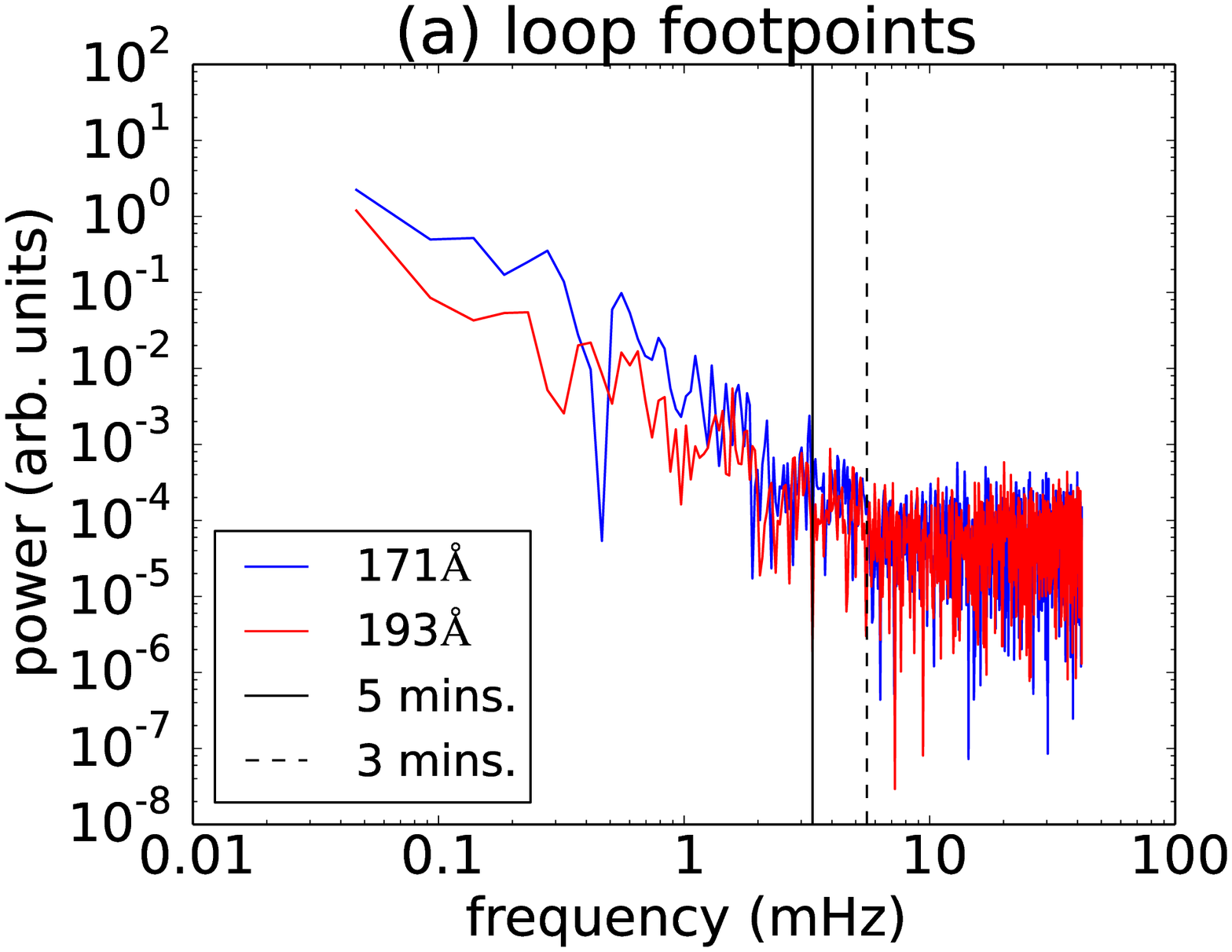}{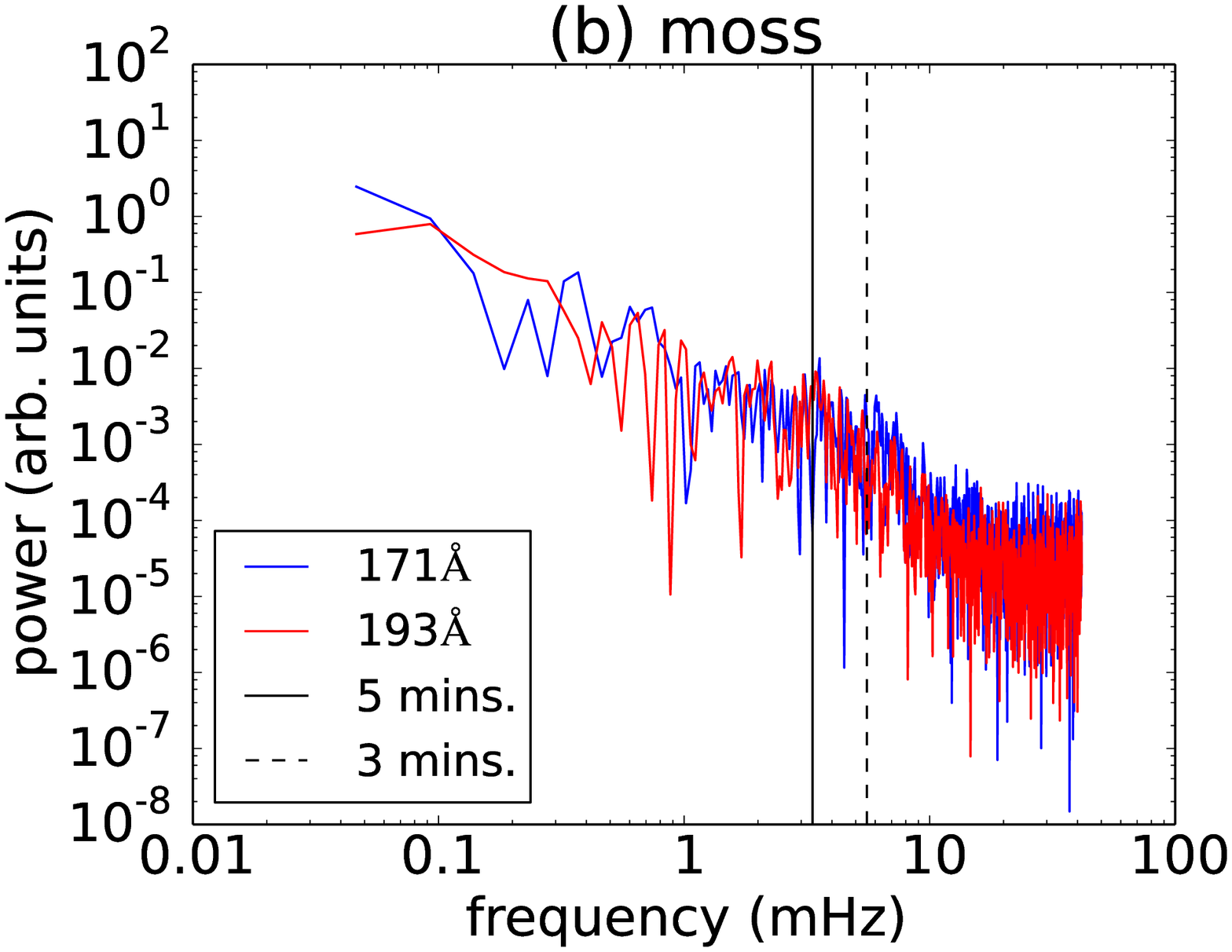}
}
\centerline{
\plottwo{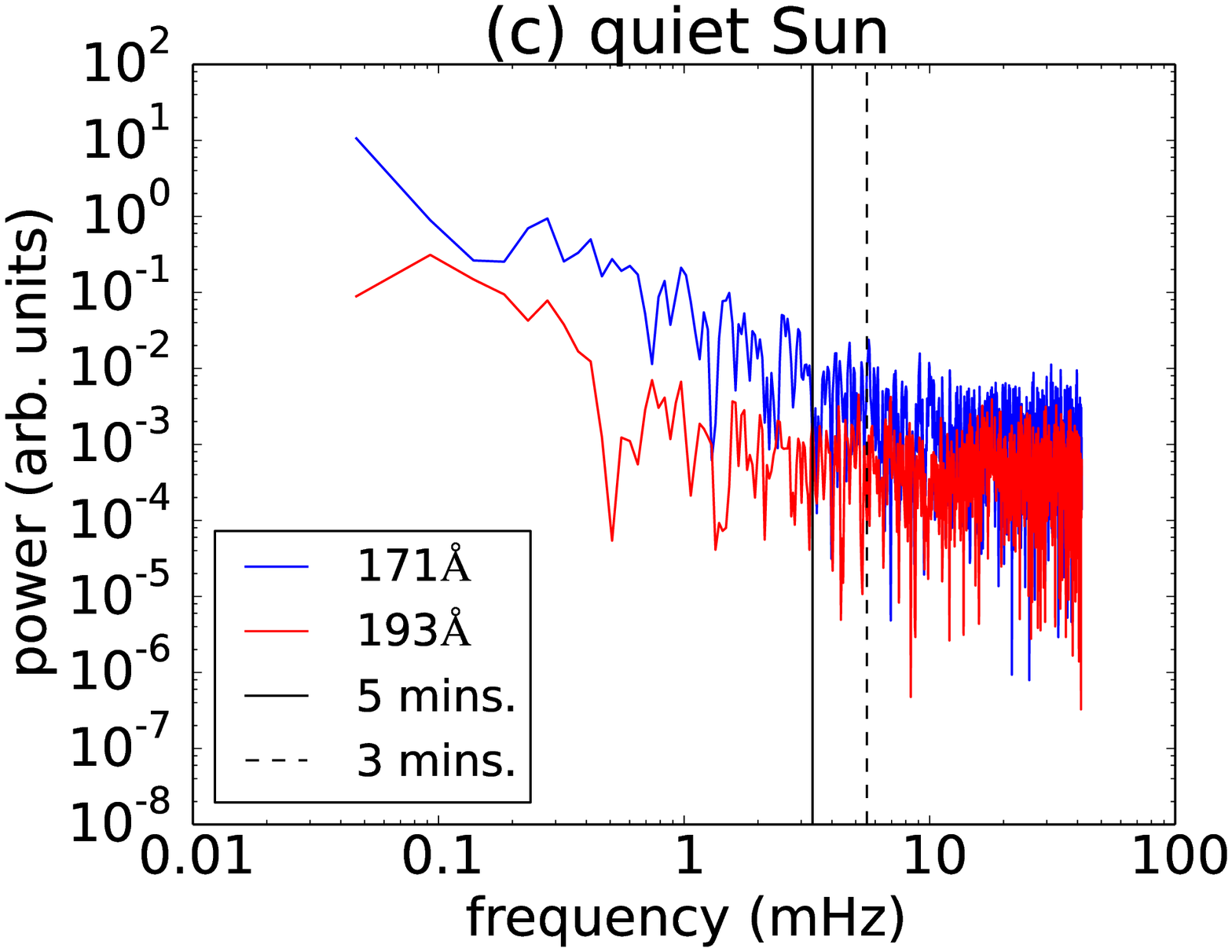}{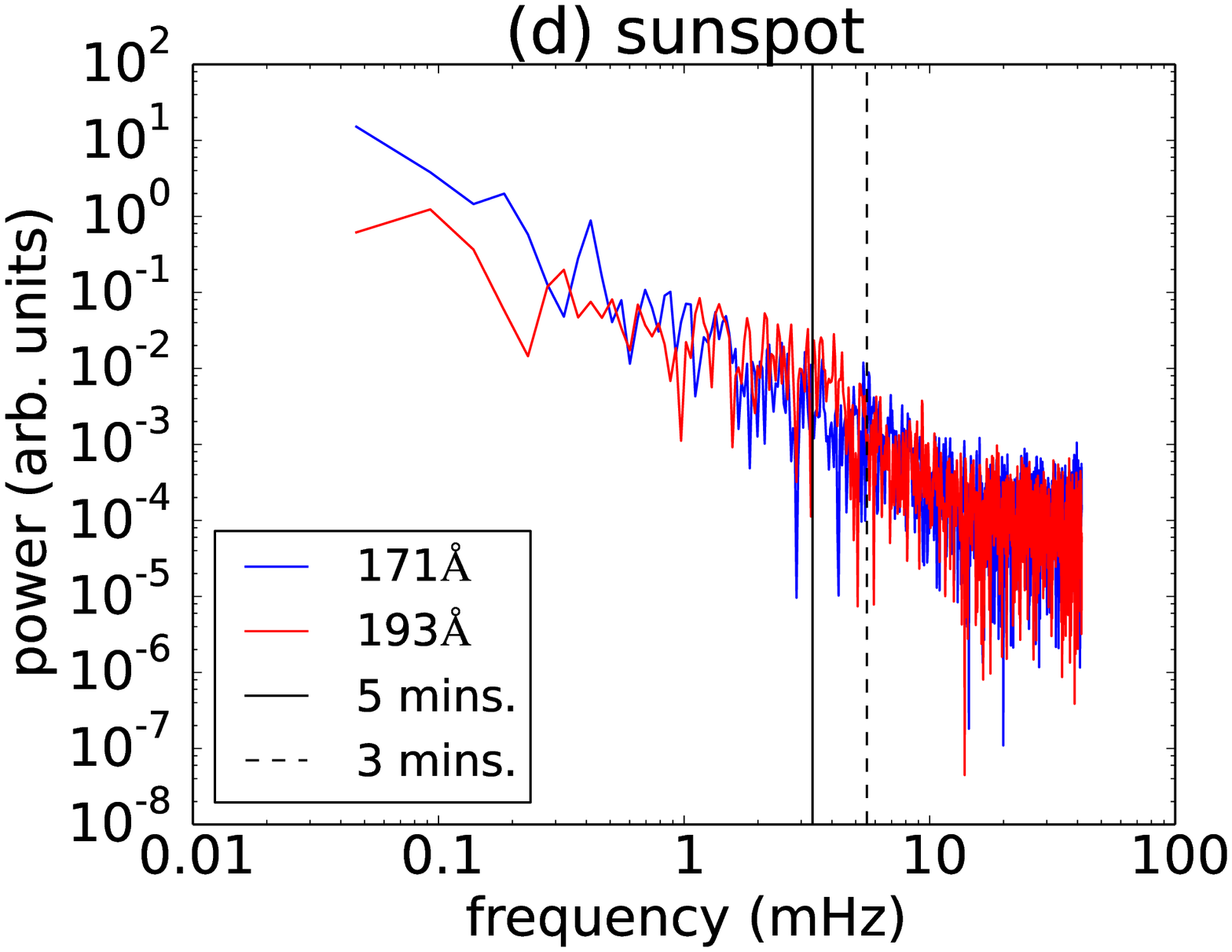}
}
\caption{Log-log plot of the Fourier power of a time-series at the AIA
  pixel overlying a randomly chosen pixel in each of the regions
  considered (see Figure \protect\ref{fig:loc171193}).  The vertical
  black lines indicate the 3 and 5 minute frequencies.}
\label{fig:compare171193}
\end{figure}

Consider the emission from a single pixel, $I(t)$.  The Fourier power
spectrum for this emission was calculated as follows.  First, the
time-series was normalized by calculating $(I(t) - \langle I(t)
\rangle)/\langle I(t) \rangle$. Secondly, the normalized time-series
was apodized using the Hanning window, reducing ringing effects in the
\PS.  The full 12 second cadence is used, yielding time series of 1800
samples.  The full spatial resolution is also retained in order to
obtain fine detail on the variation of frequency content as a function
of location.  Figure \ref{fig:compare171193} show example power
spectra, plotted using a log-log scale, from a single location inside
each of the regions shown in Figure \ref{fig:loc171193}.  The shape of
the power spectra indicate that the Fourier power approximately
follows a \PL\ at lower frequencies, and flattens out at higher
frequencies.

There is evidence that these \PLs\ extend to much longer
frequencies than those considered here, down to approximately 0.01
mHz. \cite{2014AA...563A...8A} integrate emission over small portions
of active regions and the quiet Sun as observed in the 195\AA\ passband
images from EIT and show that the resulting time-series has an
approximate \PL\ \Fps\ over the frequency range 0.01 - 1 mHz.

\begin{figure}
\centerline{
\epsscale{0.8}\plottwo{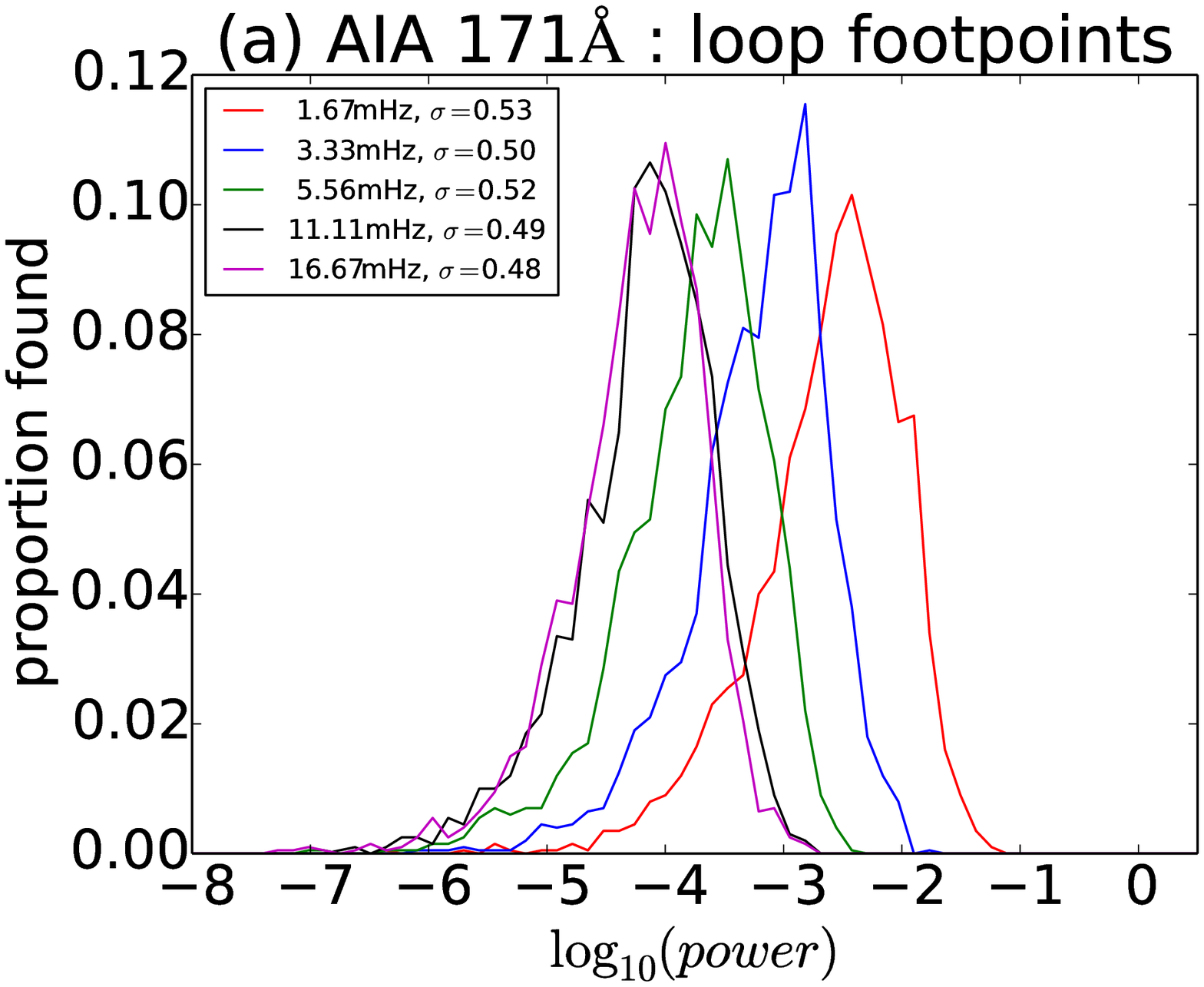}{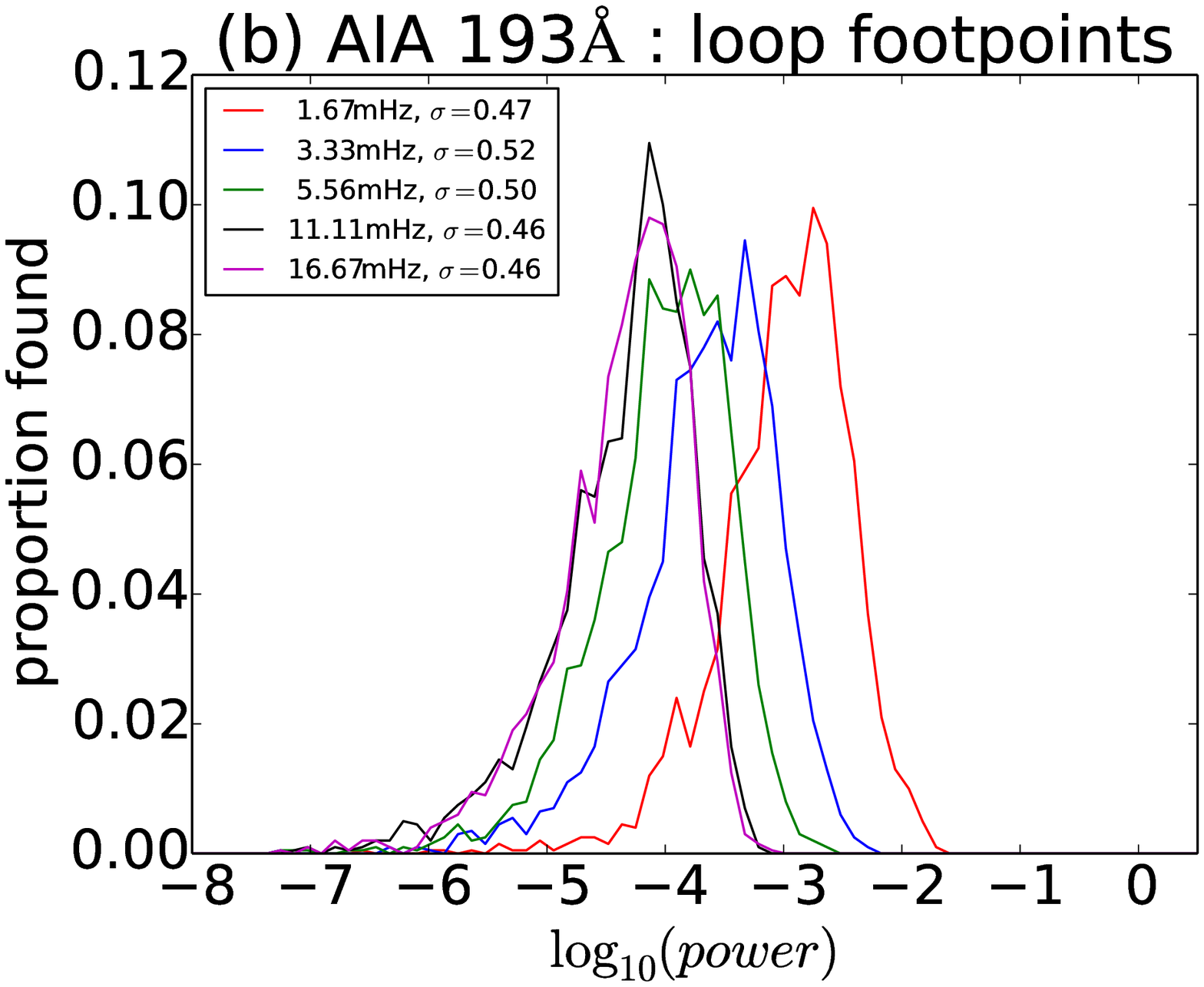}
}
\centerline{
\epsscale{0.8}\plottwo{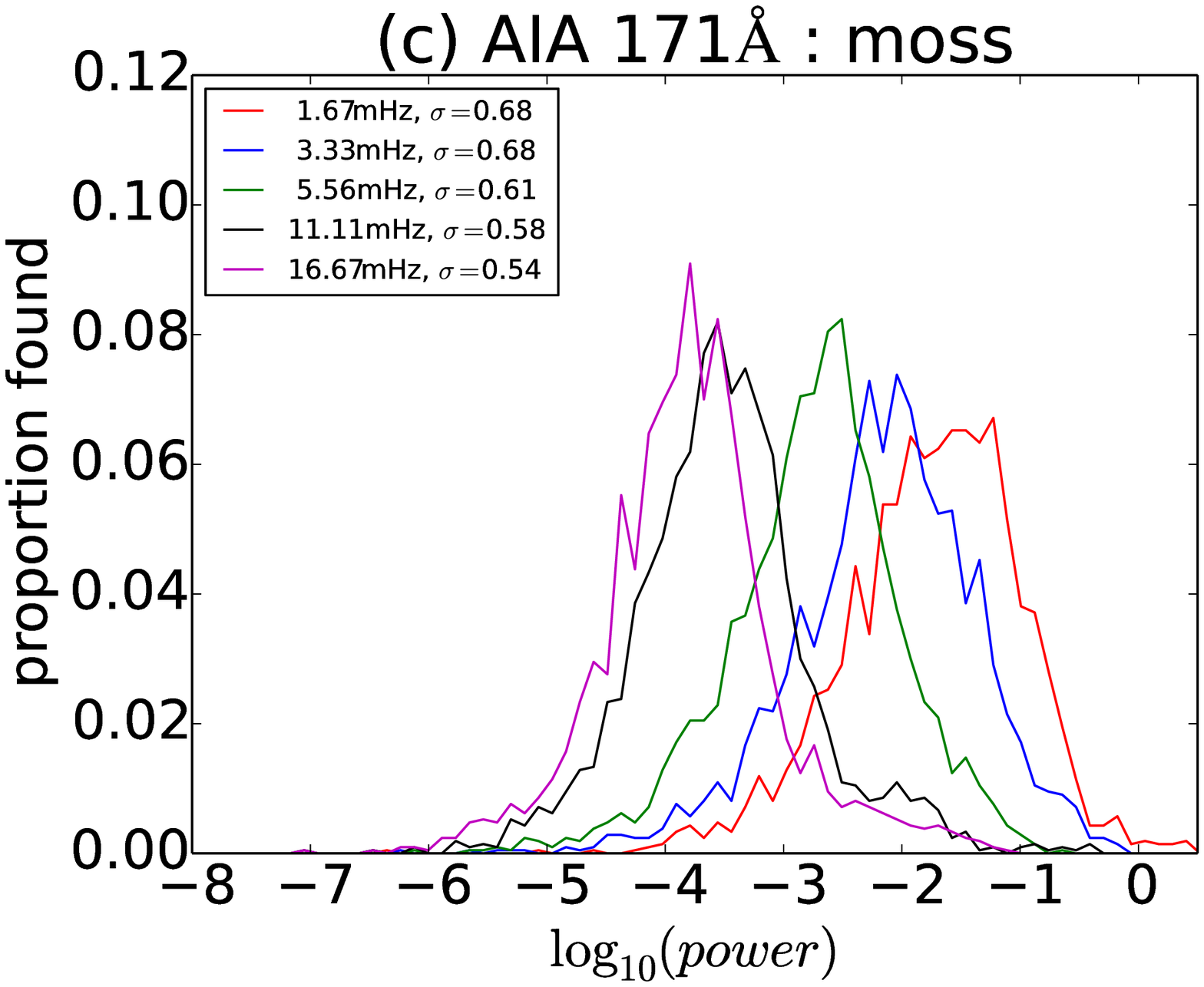}{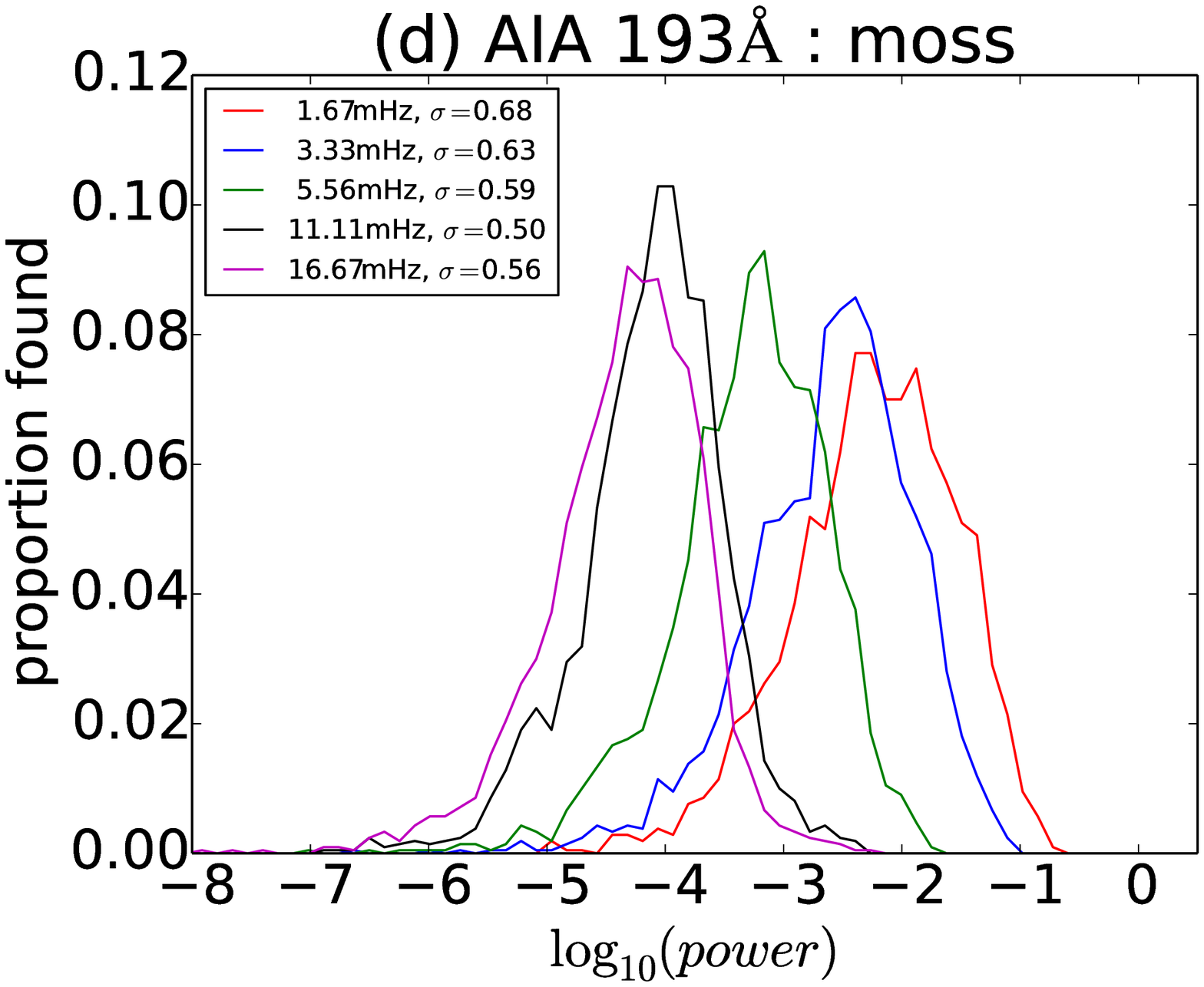}
}
\centerline{
\epsscale{0.8}\plottwo{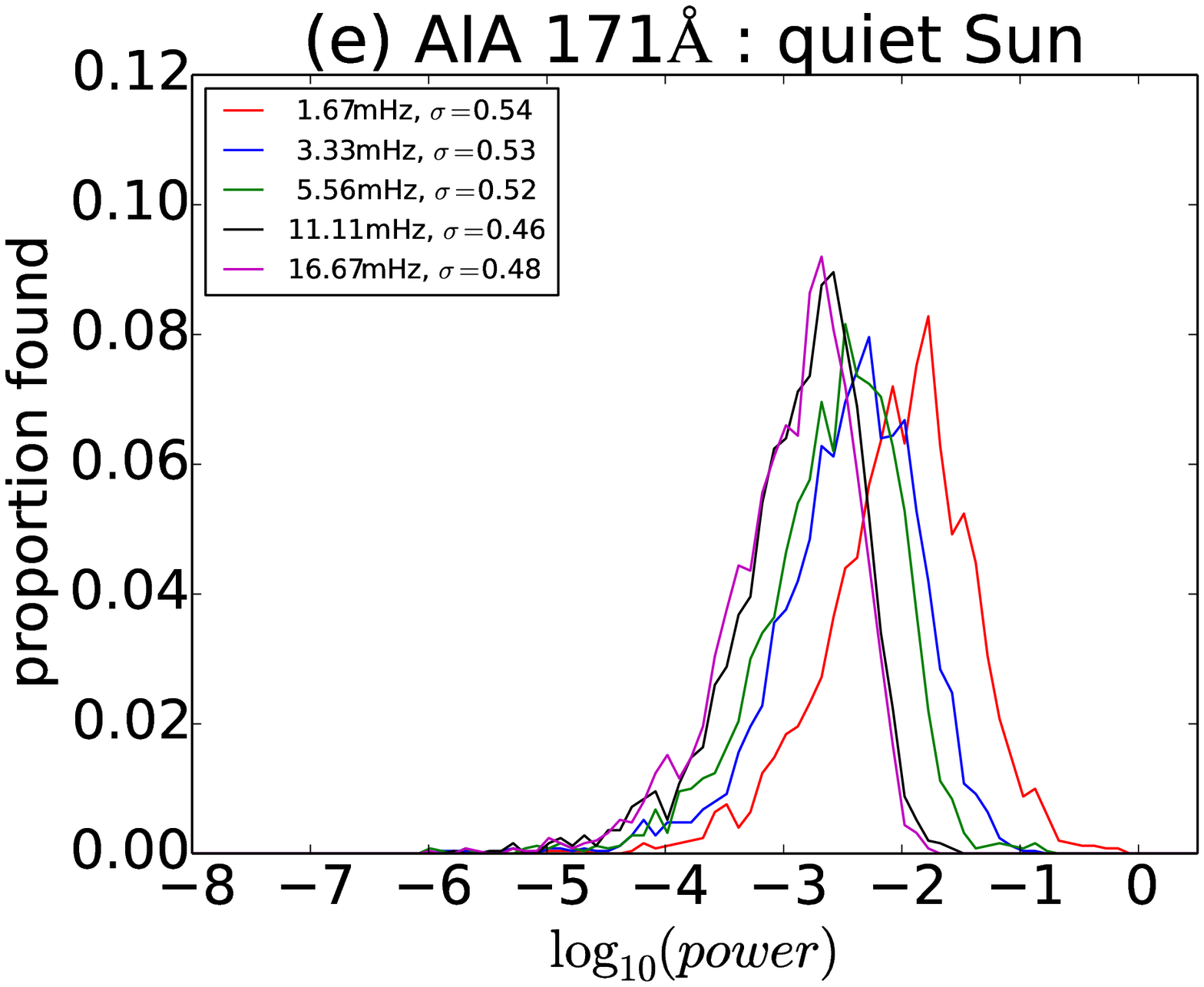}{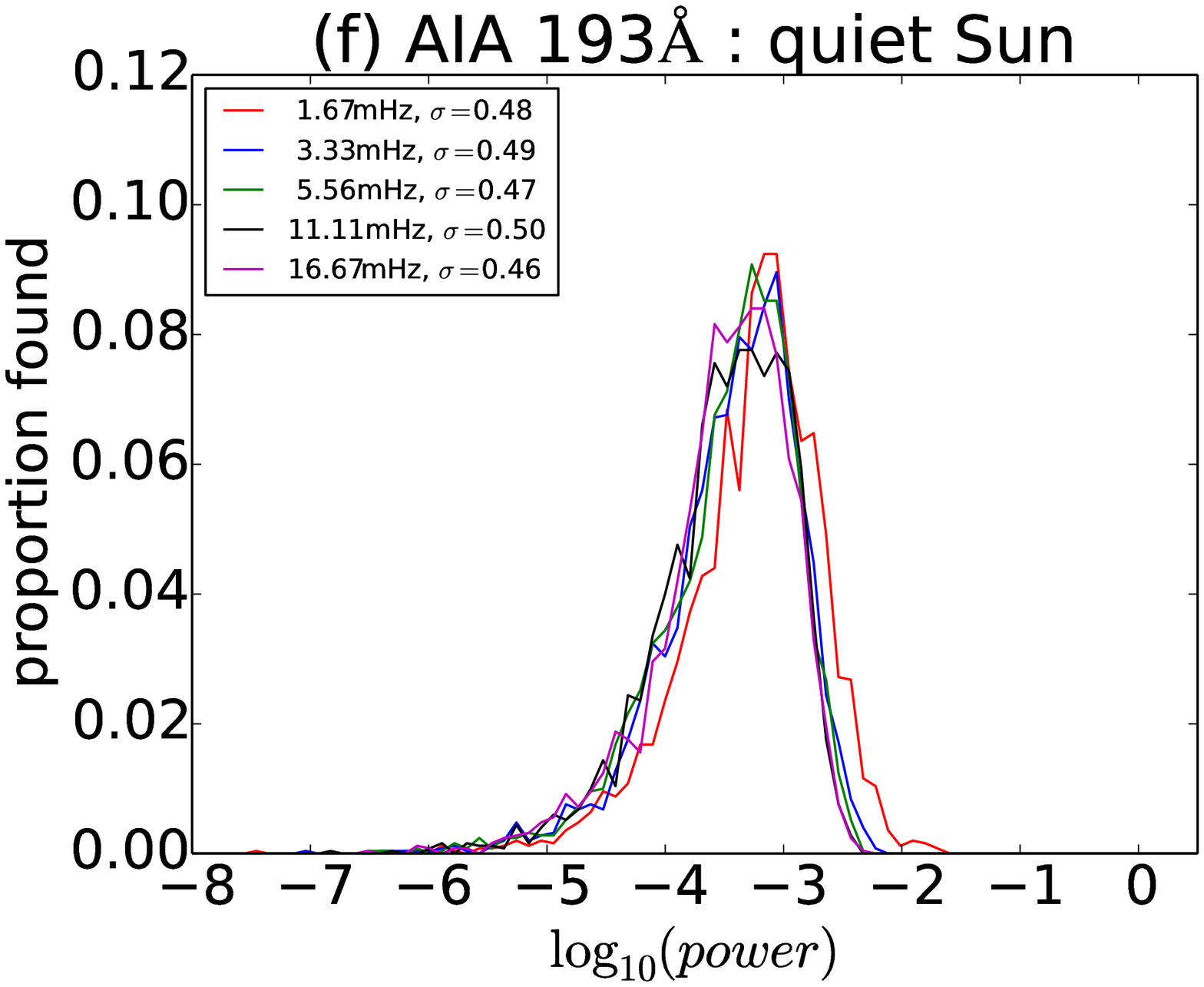}
}
\centerline{
\epsscale{0.8}\plottwo{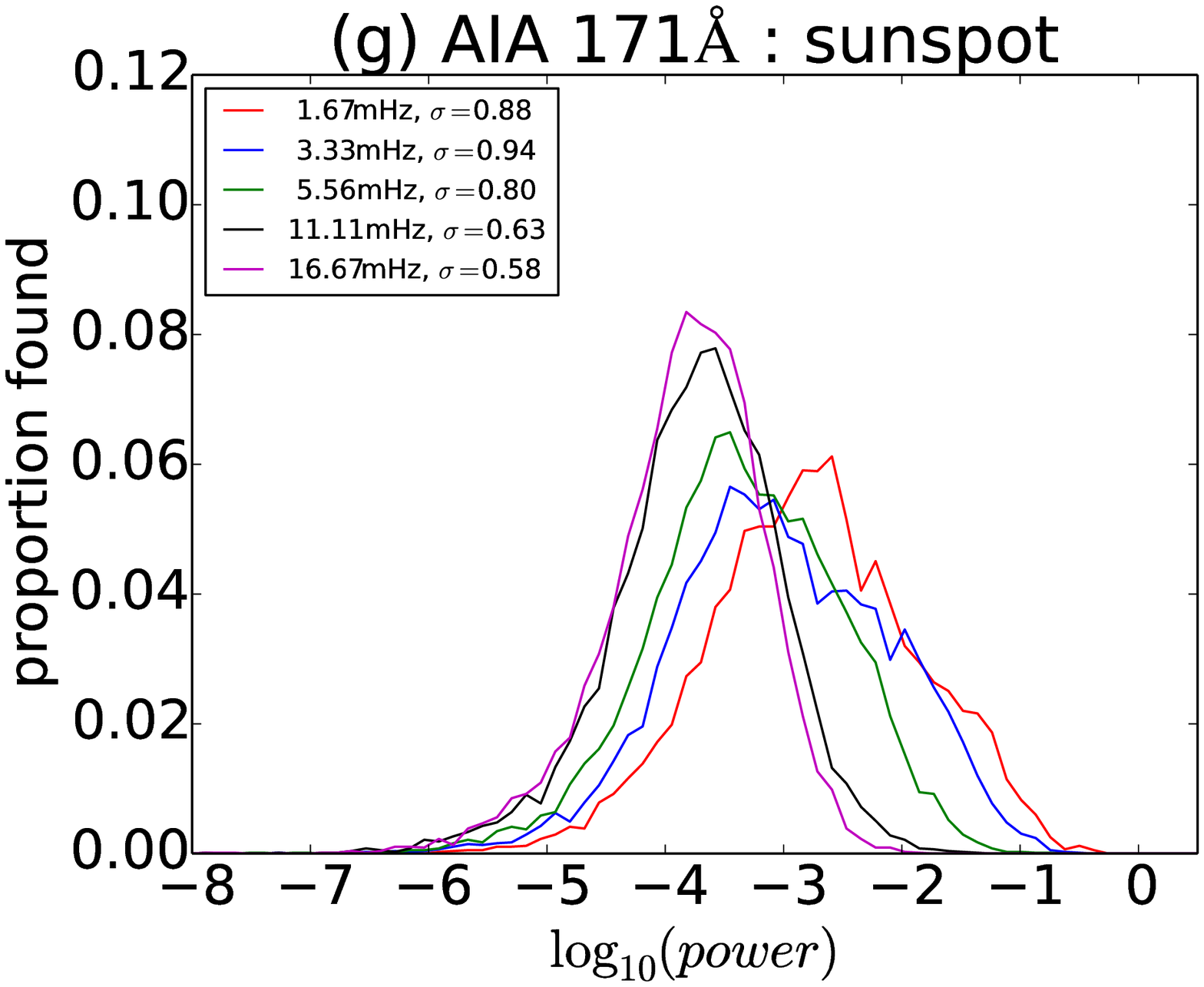}{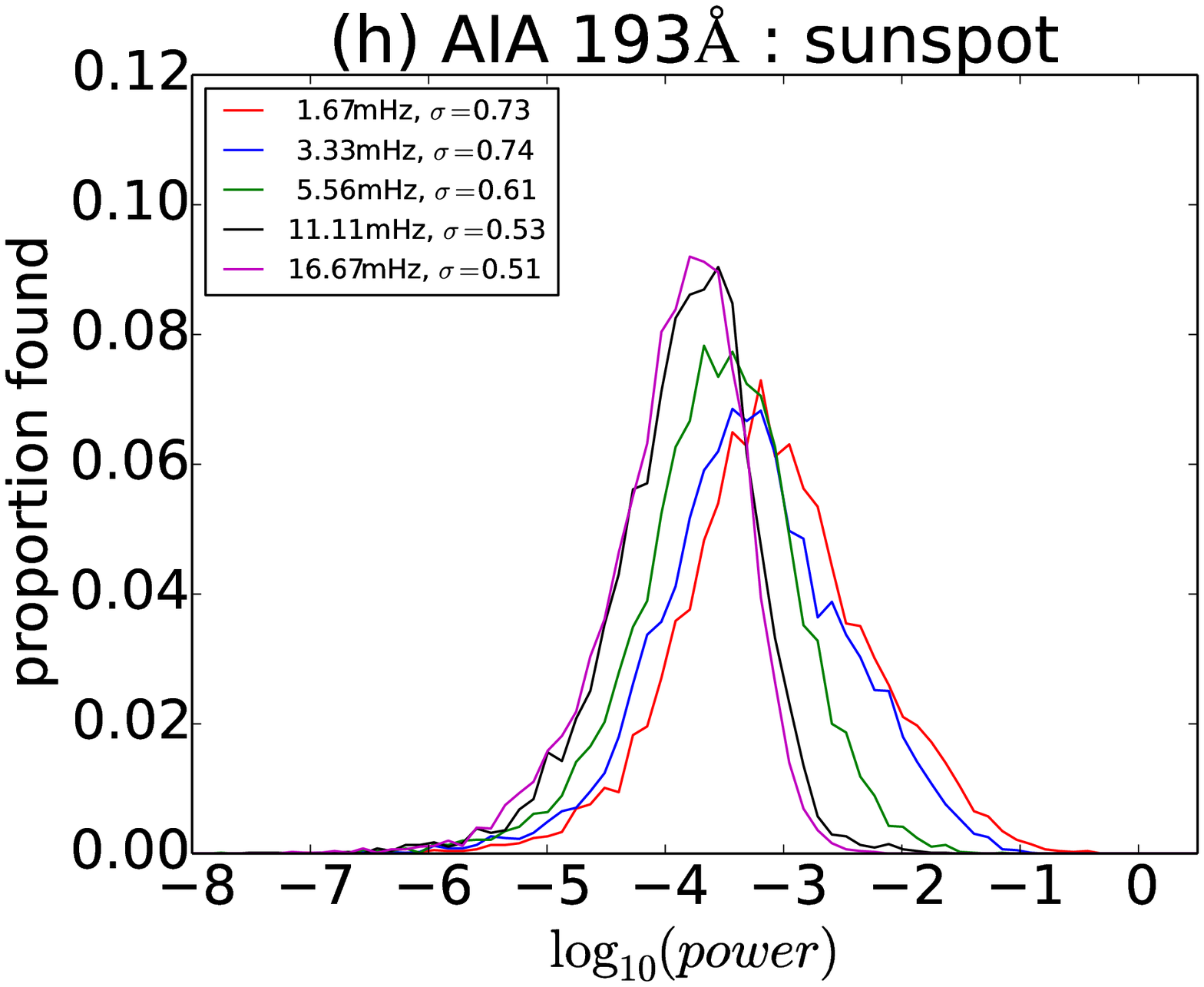}
}
\caption{{\BF The proportion of the number of pixels in each image
    that have a given Fourier power at some sample Fourier
    frequencies.  The left hand column gives results for AIA 171\AA,
    and the right hand column, AIA 193\AA.  Regions are indicated in
    each plot title.}}
\label{fig:dist171193}
\end{figure}
Figure \ref{fig:dist171193} shows the probability distribution of the
Fourier power of the pixels in each region at five selected
frequencies, for all the pixels in each region.  These distributions
are centrally peaked, show a slightly heavier tail towards lower
Fourier powers (negative skew), are slightly broader when compared to
a Gaussian distribution, and are all unimodal.  These observations
guide the choice of how to average the \PS\ over each of the selected
region.  The nature of these distributions suggest that the arithmetic
mean of the Fourier power is biased by the very largest powers.
Therefore, the mean value of the logarithm of the Fourier power is a
better description of the range of values in the data (this is
equivalent to the geometric mean of the Fourier power).  The \PA\ used
below are calculated from the mean value of the logarithm of the
Fourier power.

\begin{figure}
\centerline{
\epsscale{0.75}\plottwo{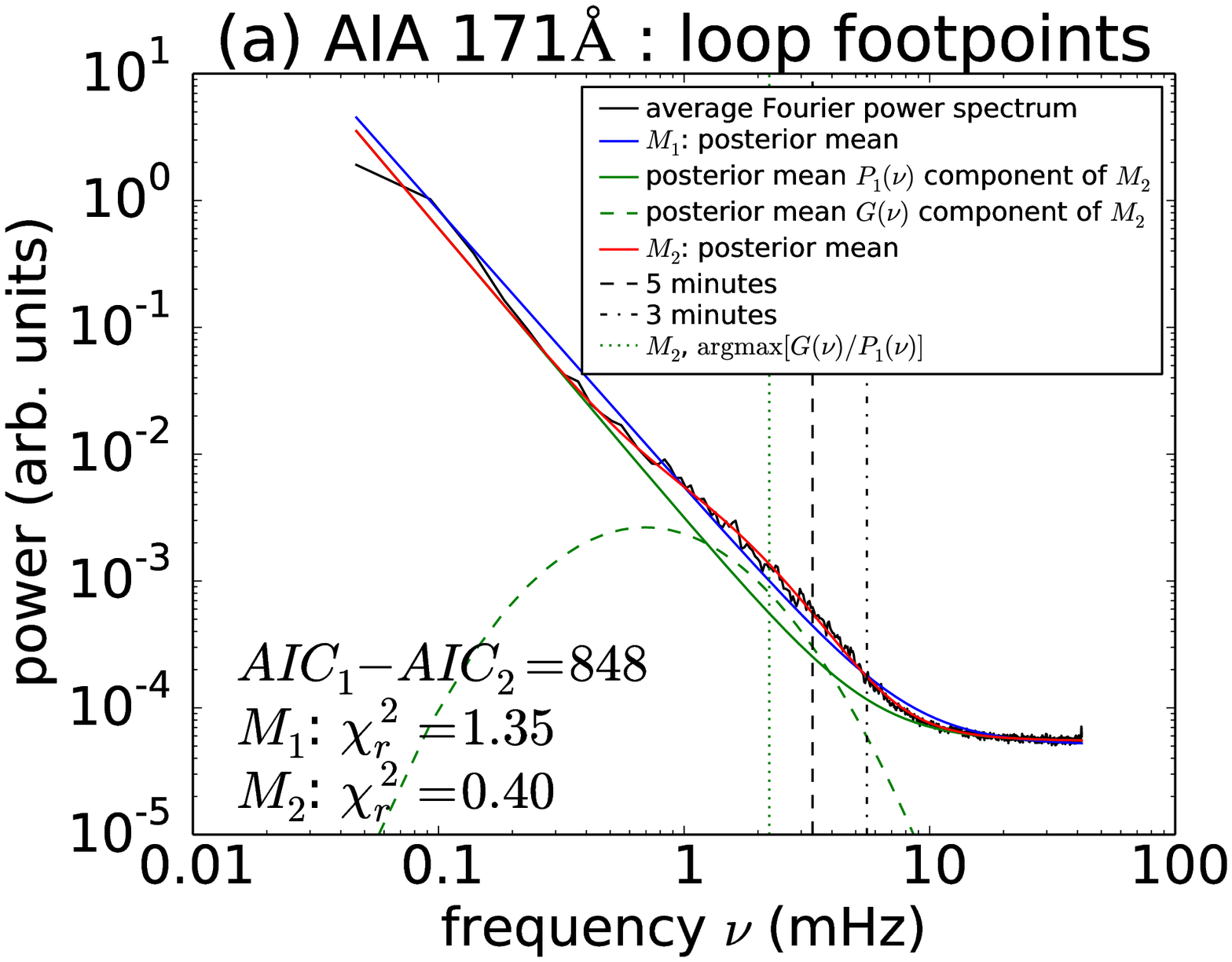}{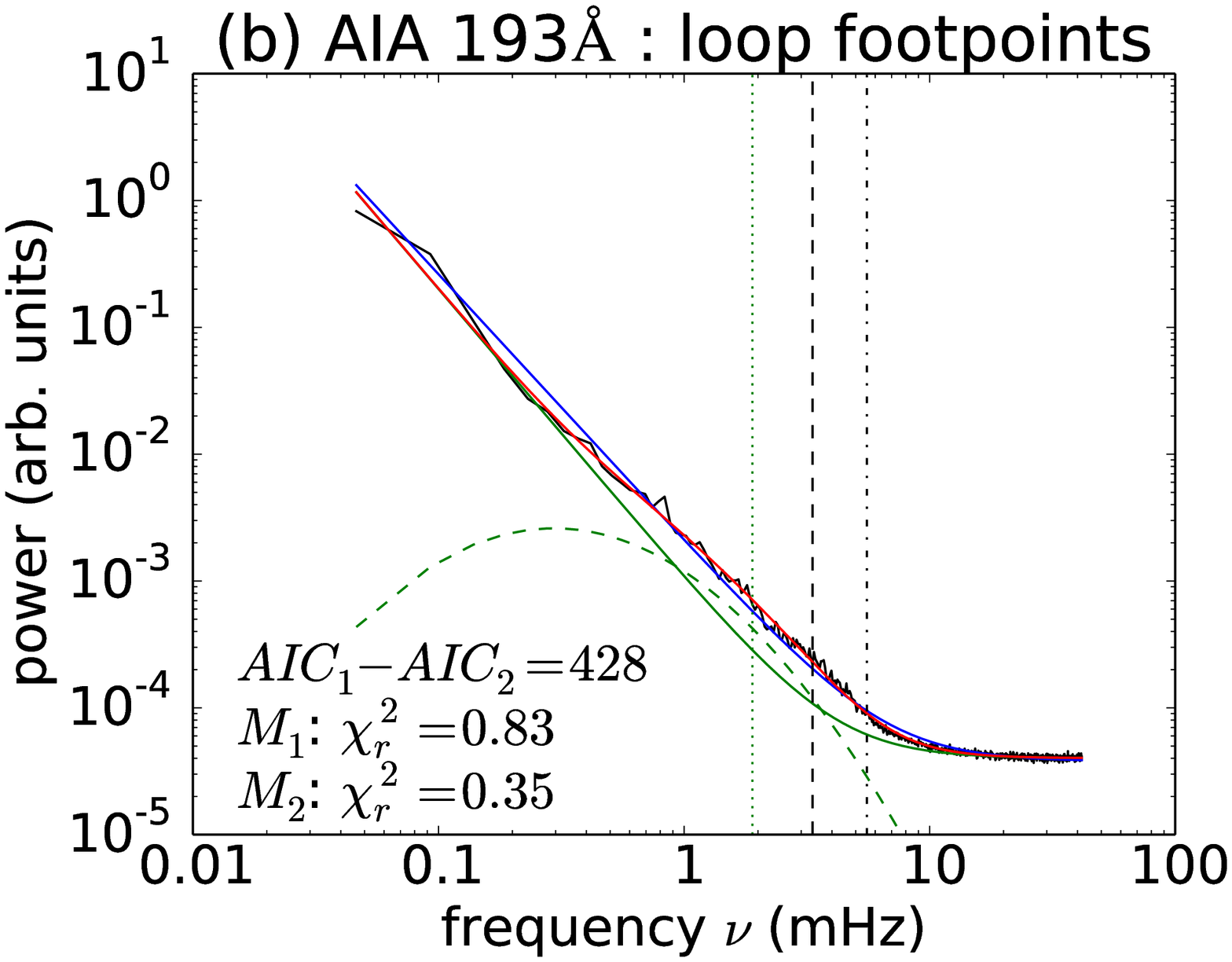}
}
\centerline{
\epsscale{0.75}\plottwo{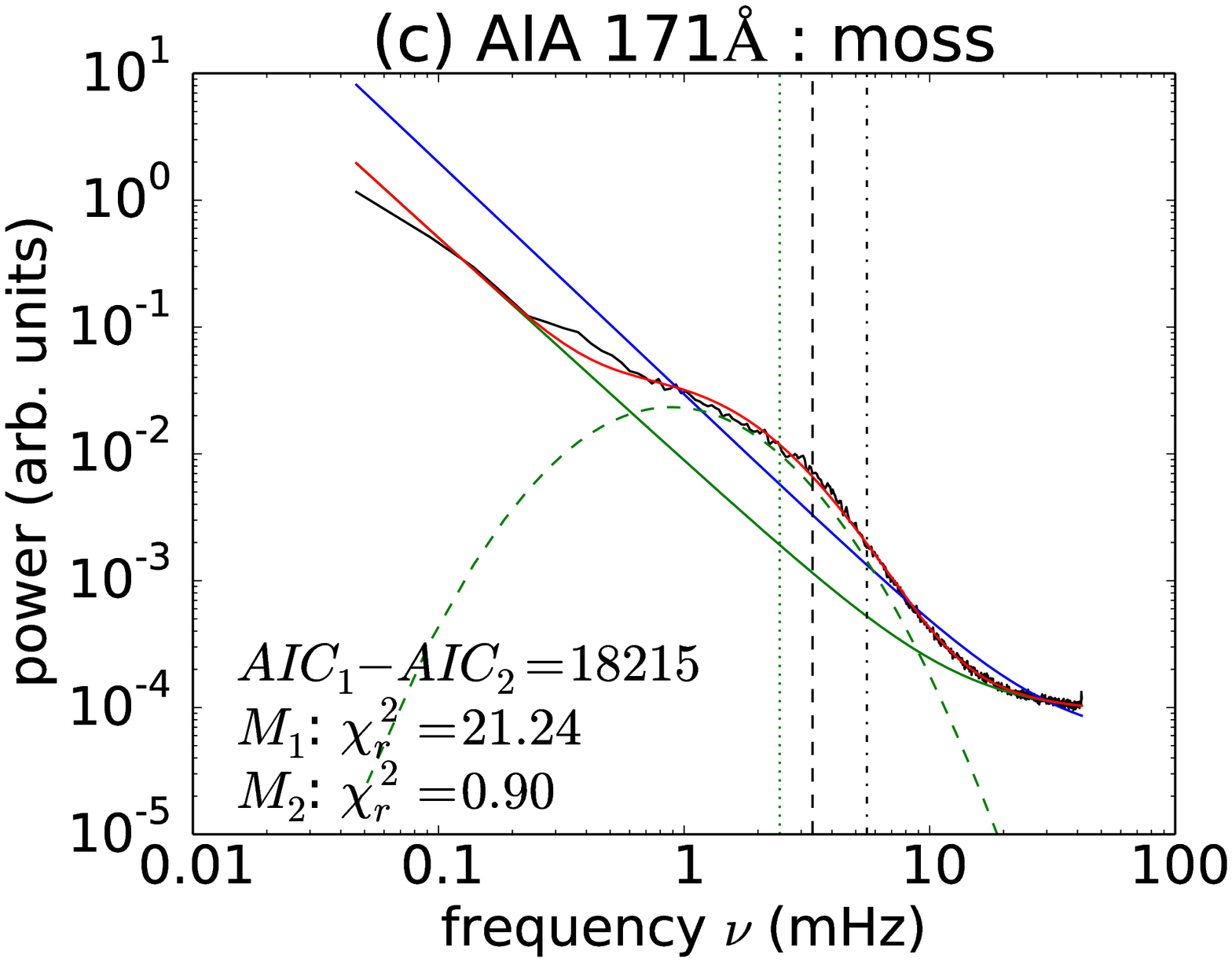}{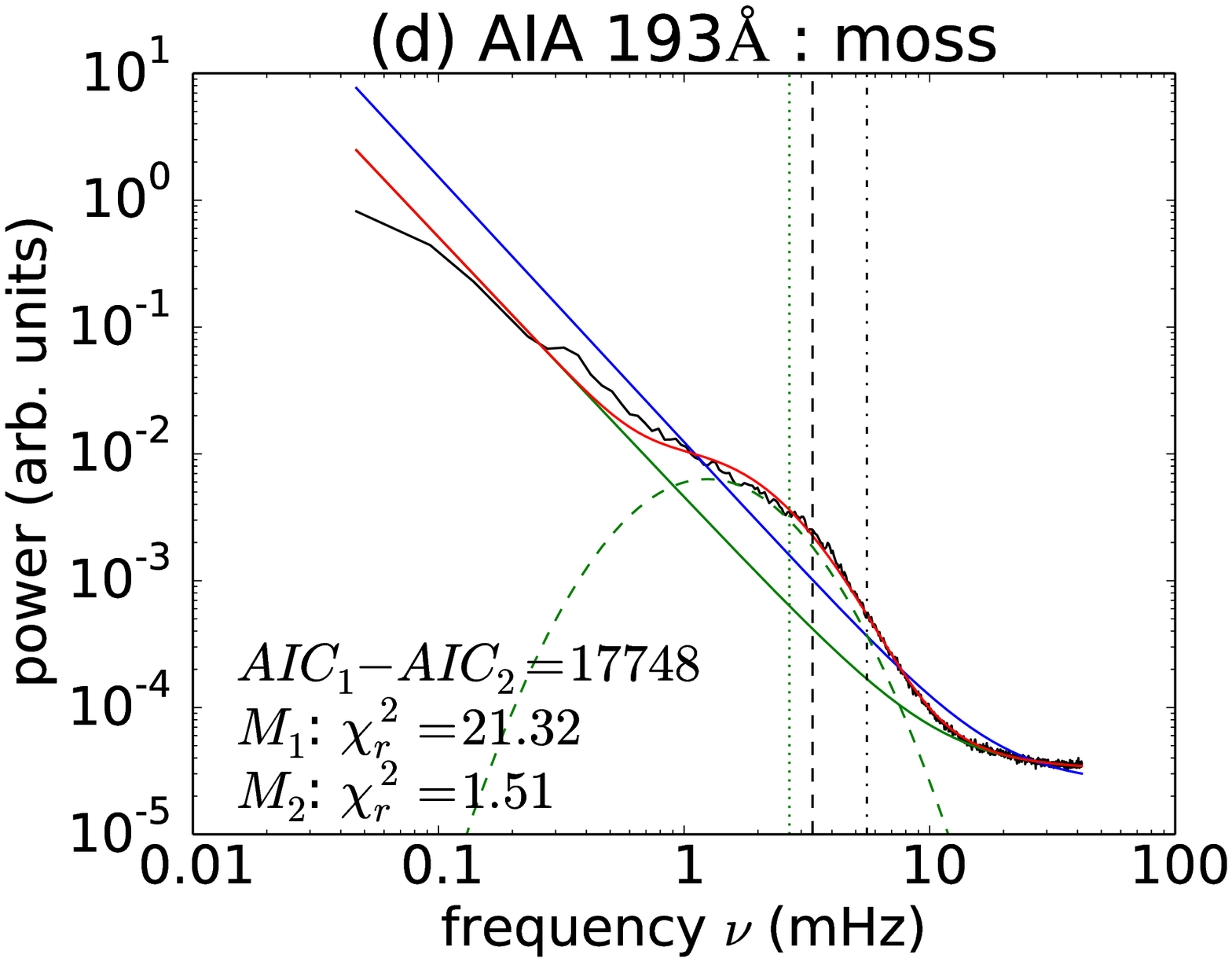}
}
\centerline{
\epsscale{0.75}\plottwo{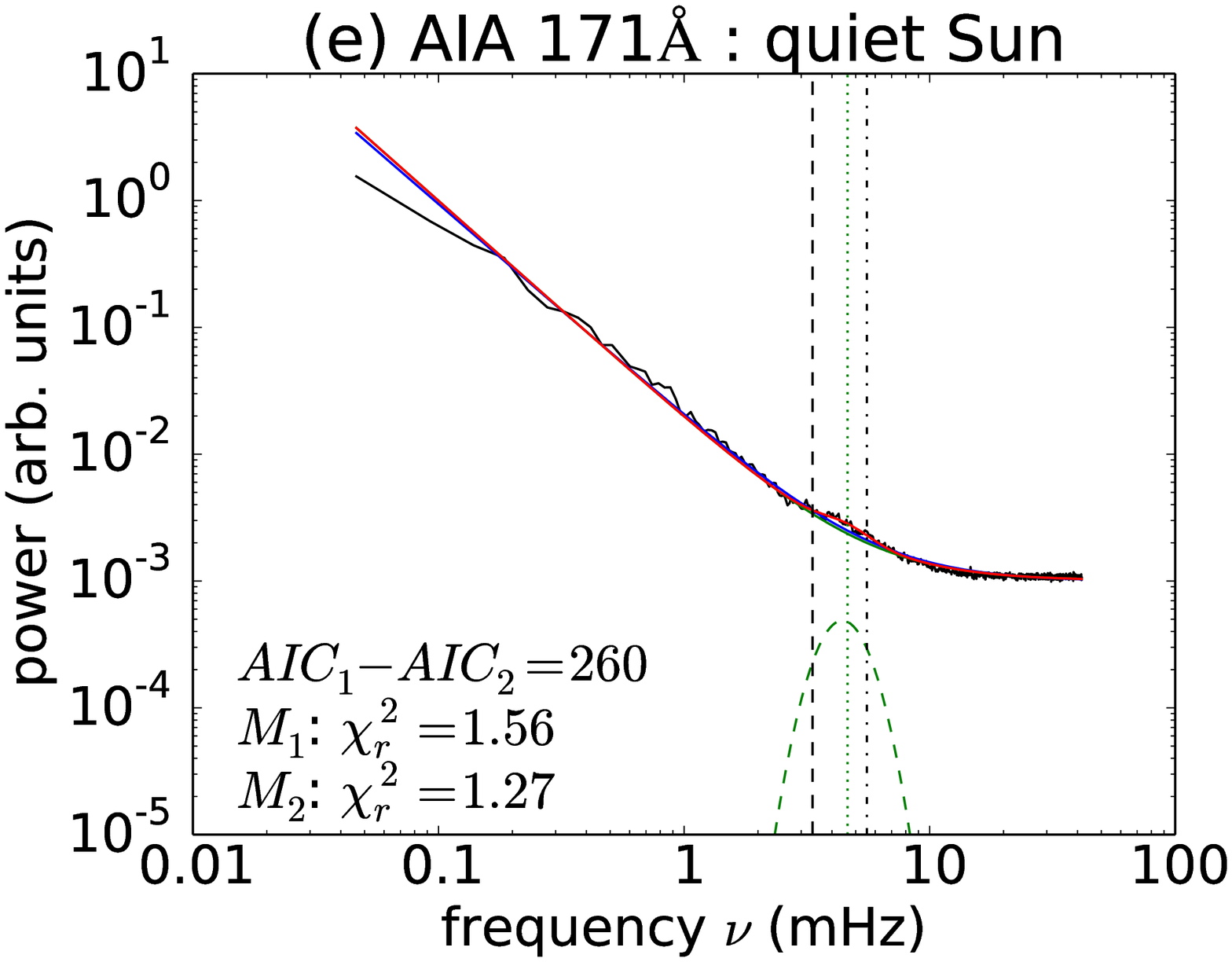}{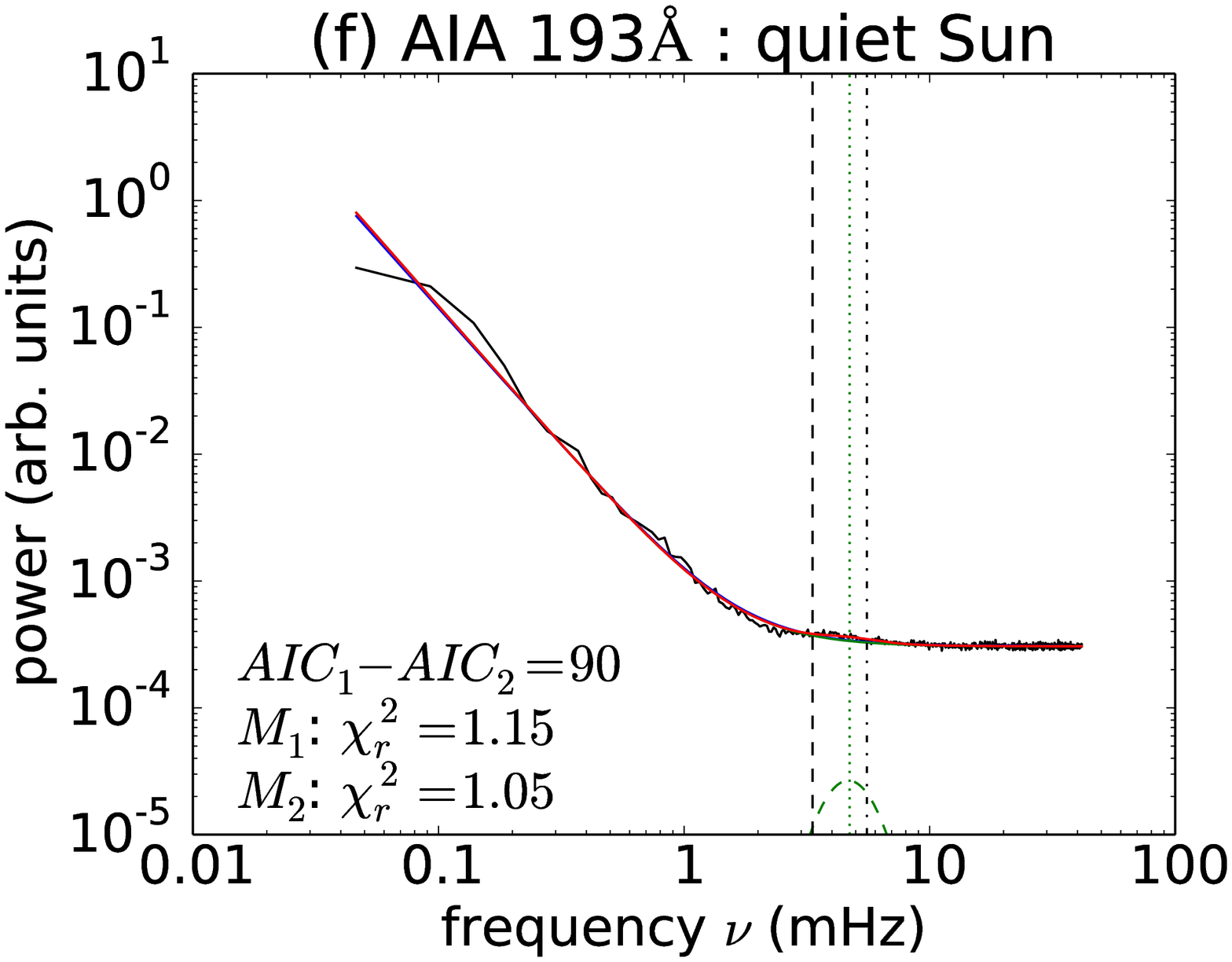}
}
\centerline{
\epsscale{0.75}\plottwo{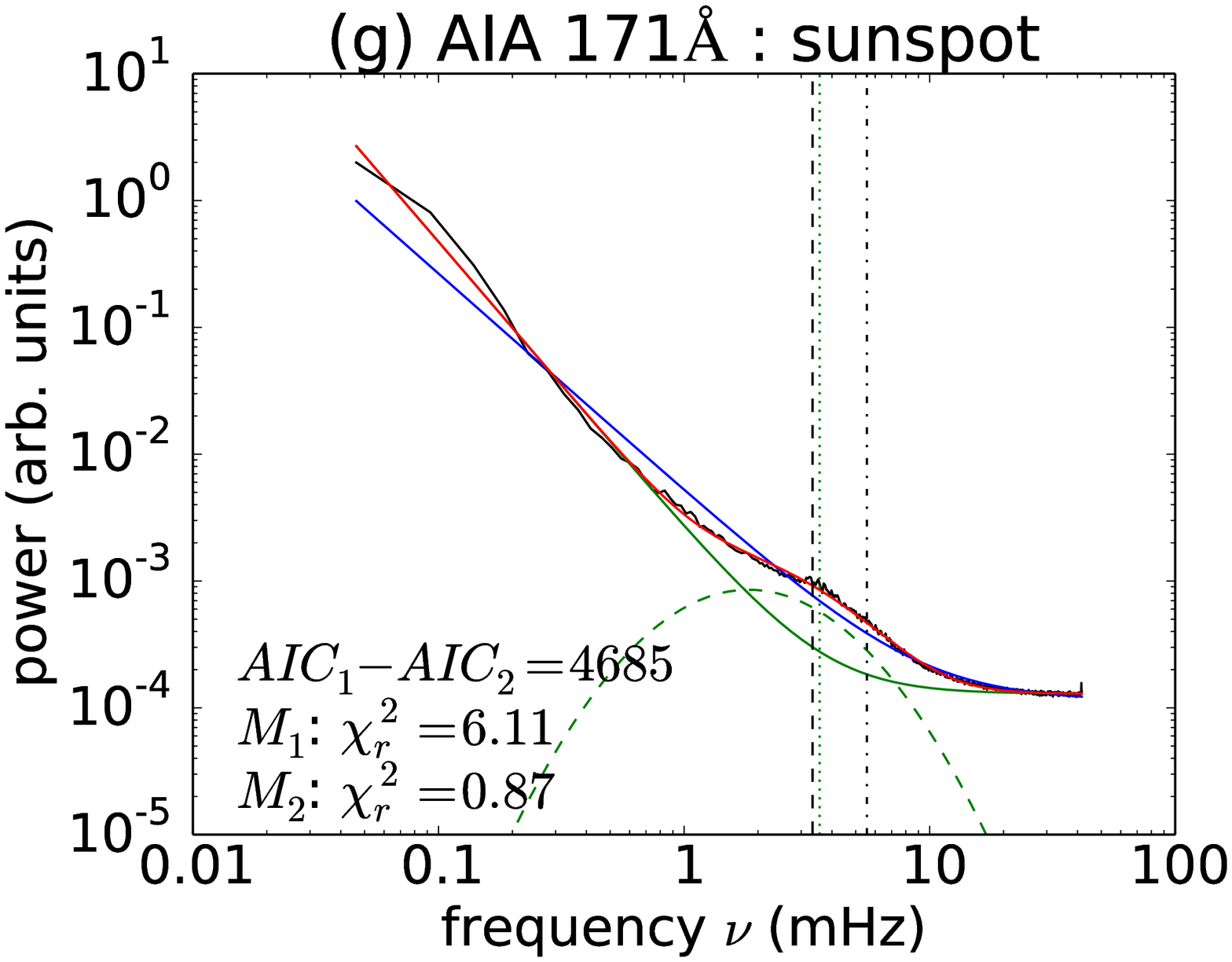}{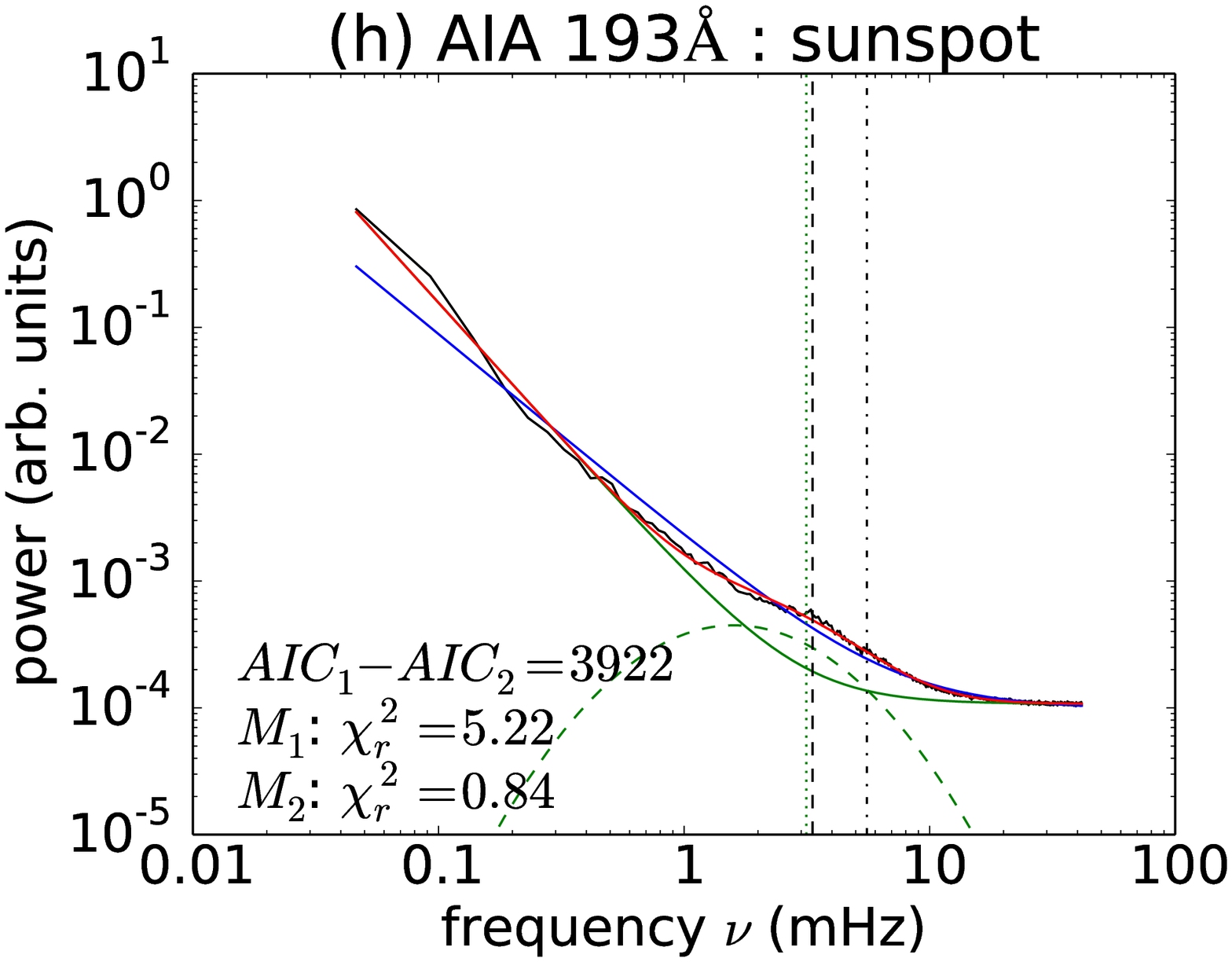}
}
\begin{singlespace}
 \caption{Geometric mean of the \protect\Fpa\ (black
    solid line) in each of the four regions studied in AIA
    171\AA\ (left column) and AIA 193\AA\ (right column).  Also shown
    are the posterior mean fits for model $M_{1}$ and $M_{2}$ (blue
    and red lines respectively; see Section \ref{sec:anal} and
    Appendix \ref{sec:app:ind} for more detail).  The AIC describing
    the preference for model over the other is also indicated in each
    plot.  The reduced $chi^{2}_{r}$ statistic for Model 1 and Model 2
    are given. The two components to Model 2, $P_{1}(\nu)$ and
    $G(\nu)$ are also shown in green; the location of the maximum of
    the $G(\nu)/P_{1}(\nu)$ ratio is indicated by the green dotted
    line.  Model fit parameter values and their 68\% credible
    intervals are quoted in Table \ref{tab:parameters}.}
\end{singlespace}
\label{fig:fit171193}

\end{figure}

The (geometric) \mFpa\ are shown in Figure \ref{fig:fit171193} for
each of the physical regions considered.  It is clear that, in
general, the \mFpa\ exhibit power-law like characteristics.  The
results for the quiet Sun, loop footpoints and sunspot regions suggest
a \PL\ \PS.  This can be modeled as
\begin{equation}
\label{eqn:pwrlaw}
\mbox{Model $M_{1}$}: P_{1}(\nu) = A\nu^{-n} + C,
\end{equation}
where $\nu$ is the frequency, $A>0$ is a proportionality constant and
$n>0$ is the \PL\ index.  The quantity $C>0$ models the high-frequency
/ low power end of the spectrum where the detection properties of the
detector apparatus are assumed to dominate the observations.

The moss results for 171\AA\ and 193\AA\ are notably different from
the other results.  In comparison to the general trend observed in
other regions, there appears to be excess Fourier power in the range
1-10 mHz.  \cite{2003ApJ...595L..63D} show that {\BF TRACE}
171\AA\ time-series of bright upper transition region emission above
active region plage (also known as moss) is a source of wavelet
power. Periods of significant wavelet power are found in the range 200
to 600 seconds, and typically persist for 4–7 cycles.  Later work by
\cite{2005ApJ...624L..61D} suggested that spicule flux tubes tilted
with respect to the solar surface provide a mechanism by which
oscillatory power from lower in the atmosphere may be channeled to
upper portions of the atmosphere.  This suggests that a second model
$M_{2}$ for the \Fps\ in regions should be considered.  The second
model $M_{2}$ has two contributions to the overall \Fps; a background
power-law of the form of $P_{1}$, and a contribution that is more
localized, $G(\nu)$.  The model is given by
\begin{equation}
\label{eqn:pwrlawbump}
\mbox{Model $M_{2}$}: P_{2}(\nu) = P_{1}(\nu) + G(\nu)
\end{equation}
where the localized contribution is described by
\begin{equation}
\label{eqn:bump}
G(\nu) = \alpha\exp\left[-\frac{(\ln(\nu)-\beta)^{2}}{2\delta^{2}}\right].
\end{equation}
Both these models are fit to the geometrically averaged Fourier power
spectrum in both wavelengths and in all four regions (Figure
\ref{fig:fit171193}).  {\BF Model fitting and parameter estimation are
  implemented using a Bayesian probability based approach.}  This
approach was chosen for its flexibility and consistent error
estimation properties \citep{2013ApJ...769...89I}.  The details of the
model fitting process are described in Appendix \ref{sec:app:ind}.
The position of $G(\nu)$ is restricted by permitting $\beta$ to vary
in the range $0.1-10$ mHz only.  This restriction, which is imposed in
the fitting process (see Appendix \ref{sec:app:ind}), arises from the
observations of significant wavelet power in the period range 200 -
600 seconds as found by \cite{2003ApJ...595L..63D}.  {\BF After this
  fitting process is complete, the models are compared to decide which
  model best describes the data.  The details of the model comparison
  process are described in the following section.}

\subsection{Results}\label{ssec:results}
Figure \ref{fig:fit171193} show the fit results of each model to the
data.  For details on the fitting procedure, see Appendix
\ref{sec:app:ind}.
{\BF The two models are assessed for how well they describe the data in two
different ways. Model selection is implemented using the Akaike Information Criterion
(AIC; \citealp*{akaike}).  The AIC measures the amount of information
lost when fitting a model to data \citep{2000Ap&SS.271..213T}.  The
AIC is defined as
\begin{equation}\label{eqn:aic}
AIC \equiv 2k -2 \ln L_{max}
\end{equation}
where $L_{max}$ is the maximum likelihood achievable by the model, and
$k$ is the number of parameters in the model ($k=3$ for Model 1, and
$k=6$ for Model 2).  Model selection is guided by calculating
$\triangle AIC = AIC_{1} - AIC_{2}$.  Positive values of $\triangle
AIC$ indicate that Model 2 loses less information compared to Model 1,
leading to a preference for Model 2 over Model 1
\citep{2007MNRAS.377L..74L}.  The AIC indicates that Model 2 is
overwhelmingly preferred to Model 1 (see Figure \ref{fig:fit171193}).}
{\BF The AIC reveals which (out of a set of models) is preferred.  It
  does not give an assessment of the goodness-of-fit of each model.
  Goodness-of-fit is estimated by calculating reduced-$\chi^{2}$,
  $\chi^{2}_{r}=\sum_{j=1}^{N}\left[(P_{i}(\nu_{j})-D_{j})/\sigma_{j}^{2}\right]/(N-k-1)$,
  where $N=899$ is the number of frequencies, $P_{i}(\nu_{j})$ is the
  model power $P_{i}$ at frequency $\nu_{j}$, $D_{j}$ is the \mFpa\ at
  $\nu_{j}$, and $\sigma_{j}$ is the estimated standard deviation of
  the data $D_{j}$ (see Appendix \ref{sec:app:ind} for more details of
  the estimation of $\sigma_{j}$).  The value of $\chi^{2}_{r}$ is
  calculated for both models.  Note that the \mFpa\ are only
  approximately normally distributed, and so caution must be used in
  interpreting $\chi^{2}_{r}$. Since $N>>k$, $\chi^{2}_{r}$ is
  understood as a measure of the average scaled variance of the model
  fit to the data. Within the approximations of the analysis, the
  $\chi^{2}_{r}$ measures show that reasonable fits are obtained for
  Model $M_{2}$ in all cases, excepting the loop footpoints.  The
  relatively low value of $\chi^{2}_{r}$ for these locations indicate
  that probably the size of $\sigma_{i}$ are over-estimated.}

Table \ref{tab:parameters} gives the parameter estimates and 68\%
credible intervals of the parameter values for model $M_{2}$, {\BF the
  maximum value of the ratio $G(\nu)/P_{1}(\nu)$ and its location}.
The \PL\ indices for all regions and both wavebands lie in the
range 1.8 to 2.3, although there is no consistency between one
waveband and another.  For example, the 171\AA\ and 193\AA\ loop
footpoint results both have the same \PL\ index, but the other
regions do not.  However, it should be noted that there is no {\it a
  priori} reason why the \PL\ indices for a region observed in two
different wavebands should be the same.  This is because each waveband
is designed to examine different features on the Sun in broadly
different temperature ranges, and may be imaging {\BF different parts
  of the same temperature-dependent physical process.  Also,
  line-of-sight effects may be different in different wavebands (and
  locations).}  Hence, the emission observed in one waveband need not have
the same statistical properties as any other.

The ratio of the two components $G(\nu)/P_{1}(\nu)$ measures the
relative contribution of the ``excess'' emission over the
\PL\ component.  The frequency $\nu_{max}$ is the frequency at which
this ratio is a maximum, and indicates the most likely frequency at
which the signal of the $G(\nu)$ contribution is strongest.  The
ratios maximize in the range $1.90 \le \nu_{max} \le 4.72$ mHz.  This
is in the frequency range suggested by the moss observations of
\cite{2003ApJ...595L..63D}.  If Model 2 is indeed a reasonable model
of the \Fpa\ in the regions studied, then the $G(\nu)$ contribution is
consistent with previous observations.

Note that neither models $M_{1}$ nor $M_{2}$ feature any narrow-band
oscillatory signals that may be present in the regions studied.  Such
features were not modeled in this study as their contribution to the
overall signal is relatively small and localized to specific
frequencies when compared to the general shape of the power spectra
over the approximately three orders of magnitude (in frequency)
studied here.  The question of how best to determine the presence or
otherwise of narrow-band oscillatory power in \Fps\ is of importance
to the study of coronal seismology, and is addressed in Sections
\ref{ssec:corseis} and \ref{sec:oscdetect}.

\begin{deluxetable}{cccccccccc}
\tabletypesize{\scriptsize} \tablecolumns{5} \tablewidth{0pt}
\tablecaption{Parameter values and uncertainty estimates derived for
  model \protect$M_{2}$.  {\BF Uncertainties with an absolute value
    less than $0.005$ are quoted as $0.00$.} {\BF Uncertainties with
    an absolute value between 0.005 and 0.01 are rounded up to have an
    absolute value of 0.01} \label{tab:parameters}} \tablehead{
  \colhead{Region} & \colhead{Waveband} &
  \multicolumn{3}{c}{\PL\ $P_{1}(\nu)$} & \multicolumn{3}{c}{lognormal
    $G(\nu)$} & \multicolumn{2}{c}{ratio
    $P_{1}(\nu)/G(\nu)$}\\ \colhead{} & \colhead{} &
  \colhead{$\log_{10}A$} & \colhead{$n$} & \colhead{$\log_{10}C$} &
  \colhead{$\log_{10}\alpha$} & \colhead{$\beta$ (mHz)} &
  \colhead{$\delta$\tablenotemark{1}} & \colhead{$\max$} &
  \colhead{$\arg\max$ (mHz)} \\ } \startdata loop footpoints & 171\AA
& $0.55^{+0.02}_{-0.02}$ & $2.29^{+0.02}_{-0.02}$ &
$-4.26^{+0.00}_{-0.00}$ & $-2.30^{+0.09}_{-0.08}$ &
$0.69^{+0.09}_{-0.08}$ & $0.32^{+0.01}_{-0.01}$ & 1.45 & 2.22 \\ &
193\AA & $0.06^{+0.03}_{-0.02}$ & $2.28^{+0.04}_{-0.05}$ &
$-4.40^{+0.00}_{-0.00}$ & $-2.18^{+0.16}_{-0.22}$ &
$0.28^{+0.11}_{-0.06}$ & $0.43^{+0.03}_{-0.04}$ & 1.58 & 1.90 \\ & & &
& & & & & & \\

moss            & 171\AA  & $0.29^{+0.01}_{-0.01}$  & $1.76^{+0.01}_{-0.01}$ & $-4.04^{+0.00}_{-0.00}$  & $-1.34^{+0.02}_{-0.02}$  & $0.88^{+0.02}_{-0.02}$  & $0.34^{+0.00}_{-0.00}$    & 5.19 & 2.45\\
                & 193\AA  & $0.39^{+0.01}_{-0.01}$  & $2.05^{+0.01}_{-0.01}$ & $-4.49^{+0.00}_{-0.00}$  & $-2.00^{+0.02}_{-0.02}$  & $1.24^{+0.03}_{-0.03}$  & $0.27^{+0.00}_{-0.00}$    & 4.72 & 2.68 \\
                & & & & & & &   & & \\

quiet Sun       & 171\AA  & $0.57^{+0.01}_{-0.01}$  & $1.72^{+0.01}_{-0.01}$ & $-3.00^{+0.00}_{-0.00}$  & $-3.55^{+0.04}_{-0.04}$  & $4.41^{+0.08}_{-0.08}$  & $0.10^{+0.01}_{-0.01}$    & 0.20 & 4.63 \\
                & 193\AA  & $-0.10^{+0.01}_{-0.01}$  & $2.20^{+0.01}_{-0.01}$ & $-3.52^{+0.00}_{-0.00}$  & $-4.75^{+0.05}_{-0.05}$  & $4.68^{+0.18}_{-0.17}$  & $0.11^{+0.01}_{-0.01}$    & 0.08 & 4.72 \\
                & & & & & & &   & & \\

sunspot         & 171\AA  & $0.43^{+0.01}_{-0.01}$  & $2.26^{+0.02}_{-0.01}$ & $-3.89^{+0.00}_{-0.00}$  & $-2.80^{+0.02}_{-0.02}$  & $1.83^{+0.05}_{-0.05}$  & $0.32^{+0.00}_{-0.00}$    & 2.07 & 3.56 \\
                & 193\AA  & $-0.09^{+0.01}_{-0.01}$  & $2.14^{+0.02}_{-0.02}$ & $-3.97^{+0.00}_{-0.00}$  & $-3.04^{+0.03}_{-0.03}$  & $1.59^{+0.07}_{-0.07}$  & $0.35^{+0.01}_{-0.01}$   & 1.55 & 3.15 \\
\enddata
\tablenotetext{1}{Values are quoted in decades of frequency}
\end{deluxetable}

\section{Discussion}\label{sec:discuss}
The presence of a \PL\ \PS\ has implications for the detection of
narrow frequency band oscillations against such a background emission,
for both non-automated and automated methods.  These are discussed in
Sections \ref{ssec:corseis} and \ref{sec:oscdetect}.  Further, the
\PL\ Fourier spectrum in coronal emission poses questions about how
that emission is formed.  Section \ref{ssec:excess} discusses possible
reasons for the localized contribution component $G(\nu)$.  A
hypothesis regarding the formation of this \PL\ spectrum is
presented in Section \ref{ssec:nplps}.

\subsection{Effect of background \PS\ assumptions on the
  detection of oscillatory power}
\label{ssec:corseis}

Section \ref{sec:anal} demonstrates the presence of an approximate
\PL\ \PS\ signal in two of the most commonly used wavebands for
coronal seismology, SDO/AIA 171\AA\ and 193\AA.  The observed spectra
generally show a \PL-like behavior at time-scales of interest in
coronal seismology, and longer.  Within the confines of the models we
have chosen, coronal moss regions show the largest contribution from a
non \PL\ \PS\ source over a relatively limited range of frequencies.
The frequencies at which the lognormal contribution is larger than the
underlying power-law distribution could be used to define an average
frequency range within which the majority contribution to the overall
Fourier power is not associated with the underlying power-law
distribution.  However, the \Fps\ in this range would still have to be
analyzed using both the power-law and lognormal contributions to the
overall power, and better estimates to the power-law parameters will
be obtained if the full frequency range of the observations are used.

The presence of a non \PL\ contribution to the \Fps\ does not change
the fact that the background emission is essentially power-law like.
For the frequencies that have been of interest in coronal seismology,
the background \PS\ is therefore scale free on average.  If there is
no time-scale that can be used to remove a background trend, then the
time-series has to be analyzed without this processing step.  However,
a lack of awareness of the background \PL\ \PS\ nature of the emission
will cause significant problems in attempting to analyze for the
presence of narrow frequency band oscillations, as the following
example demonstrates.

\begin{figure}
\epsscale{1.00}
\plotone{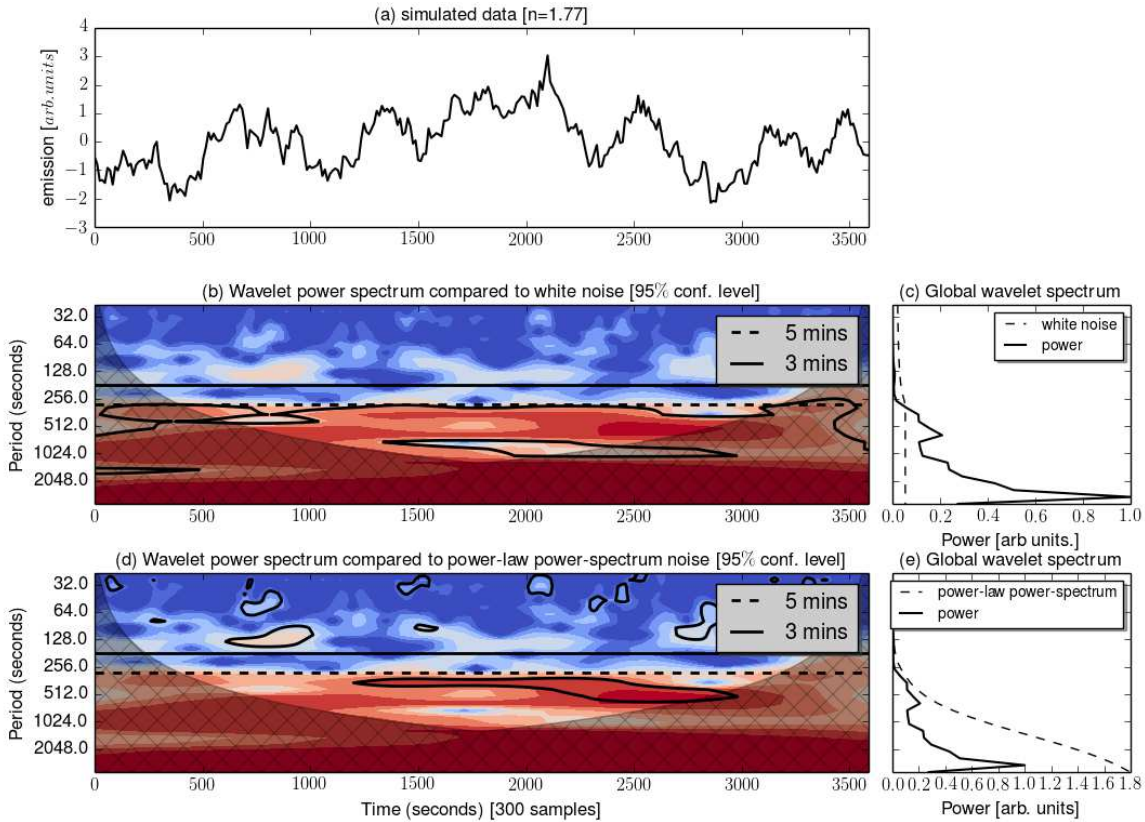}
\caption{Comparison of the effect of assuming a white or
  \PL\ \PS\ background on the detection of an oscillatory signal for
  simulated time-series.  Plot (a) shows the simulated time-series,
  generated from a \protect\PS\ \protect$P(f)\propto f^{-1.77}$ and no
  explicit oscillation included.  Plot (b) shows the wavelet power
  spectrum with the cone-of-influence (shaded area) and regions above
  the 95\% confidence level, assuming the null hypothesis that the
  time-series is purely noise, and that the noise has the same average
  power at all wavelet scales.  Plot (c) shows the global wavelet
  \protect\PS\ for this wavelet transform.  Plots (d) and (e) are the
  same as plots (b) and (c) with the null hypothesis that the
  time-series is purely noise, modeled using an auto-regressive AR(1)
  model.  Such time series assume that the current value in a
  time-series depends on its previous value plus some
  Gaussian-distributed noise (\citealp*{chatfieldtimeseries}.)  This
  time-series model gives rise to a \protect\PL\ \protect\PS\ of the
  form $\mbox{power}\propto\nu^{-n}$, where $\nu$ is the frequency of
  oscillation, and $n$ is its \PL\ index.  The code used in this
  analysis is an implementation of that described by
  \citet{1998BAMS...79...61T}.}
\label{fig:comparison}
\end{figure}

Figure \ref{fig:comparison} illustrates how a time series generated
from a random sample from a \PL\ \PS\ can be mistakenly thought to
contain a narrow-band oscillatory signal.  The time series in Figure
\ref{fig:comparison}(a) {\BF is a realization of a random process with
  a \PS\ of the form $P(f)\approx f^{-1.77}$ with no explicit
  oscillatory content, using the construction procedure detailed in
  \cite{2010MNRAS.402..307V}.  A different realization of the same
  process with the same \PS\ would almost certainly generate a
  different looking time-series, but since there is more power in
  lower frequencies, lower frequency components are more likely to be
  visible to the eye compared to higher frequency components.}  {\BF
  Figure \ref{fig:comparison}(b) shows that if the null hypothesis is
  that the time-series is purely noise, and that the noise has the
  same average power at all wavelet scales (for example, if the
  time-series were pure Gaussian noise), then the 95\% confidence
  level leads to the positive detection of significant oscillatory
  power in this time series.}  {\BF The detection claim changes if a
  different null hypothesis is used. Figure \ref{fig:comparison}(d) is
  the same as Figure \ref{fig:comparison}(b) except with the null
  hypothesis that the time-series gives rise to a \PL\ \PS\ of the
  form $\mbox{power}\propto\nu^{-n}$, where $\nu$ is the frequency of
  oscillation, and $n$ is its \PL\ index}\footnote{{\BF The model
    used to generate a \PL\ \PS\ is an auto-regressive AR(1) model.
    Such time-series assume that the current value in a time-series
    depends on its previous value plus some Gaussian-distributed noise
    \citep{chatfieldtimeseries}.  The null hypothesis that generates
    the \PL\ \PS\ is not important in order to illustrate its effect
    on the detection claim.}}.  This null hypothesis assumption, with
the same confidence level, significantly reduces the area in the
wavelet transform for which a detection may be claimed.  Even with
this significantly different background assumption, there is still a
significant amount of wavelet power above the 95\% confidence level.
However, since the data is simulated without an explicit oscillatory
component, the high wavelet power must have arisen by chance.  The
simulated data, along with the results of Section \ref{sec:anal},
suggest that when examining the wavelet transforms of time-series for
wave packets, a background \PL\ \PS\ should be assumed, along with
higher confidence levels (99\% or higher), in order to minimize the
effects of mistakenly identifying random variations in the background
\PS\ as evidence for an oscillatory signal.  {\BF Since there is a
  \PL\ \PS\ in time-series of coronal emission, evidence for the
  existence of a process that generates narrow-band oscillatory power
  has to {\BF be} measured by its power over and above that expected by the
  \PL\ \PS.}

However, even this approach to identifying oscillatory power is not
quite complete, as it simply rejects the null hypothesis that the
\PS\ can be adequately explained by a \PL\ \PS.  Model comparison
techniques similar to that described in Section \ref{sec:anal} are
required to prefer a model with oscillatory content over one
without. The difficulty in finding narrow-band features in Fourier
power spectra against a background \PL\ \PS\ has been recognized by
\cite{2010MNRAS.402..307V} in the study of XMM-Newton observations of
highly variable Seyfert 1 galaxies.  \cite{2010MNRAS.402..307V}
presents a model fitting and comparison technique that is directly
applicable to the search for oscillatory power in coronal time-series.

\subsection{Implications for coronal seismology and automated
  oscillation detection algorithms}\label{sec:oscdetect}

The discussion of Section \ref{ssec:corseis} shows that \PL\ power
spectra have a powerful effect on the claims of a detection of
narrow-frequency band oscillations in single time-series.  When
looking at extended regions of coronal imaging data, evidence for the
presence of a propagating wave in the corona is strengthened if it can
be shown that narrow frequency-band oscillations occur significantly
above the background signal in several neighboring pixels.  Spatial
extent is the key feature to claim of a detected wave.  Coronal
seismology derives its observational evidence from such detections,
and automated detection algorithms hold out the hope of greatly
increasing the number of oscillating regions detected.

Figure \ref{fig:compare171193} shows that time series from individual
pixels have power-law like power spectra.  This has implications for
the design of automated oscillation detection algorithms.  Such an
algorithm must take into account the nature of the \PS.  The power-law
\PS\ itself lacks an easily definable time-scale at or below which a
background trend can be defined.  Automated detection algorithms that
rely on background trend subtraction, such as
\cite{2010SoPh..264..403I} must therefore look to other reasons to
justify a time-scale.

As has been shown above, the assumption of a Gaussian-distributed
(white) noise when a \PL\ \PS\ is actually present, coupled with too
low a confidence level, can lead to misleading results when looking
for oscillatory signals (Section \ref{ssec:corseis} and Figure
\ref{fig:comparison}).  The detection algorithms outlined in Section
\ref{sec:int} have yet to be tested assuming a background
\PL\ \PS, and so their efficacy is unknown.
\cite{2004SoPh..223....1D} and \cite{2008SoPh..248..395S} are
wavelet-based approaches, and the discussion Section
\ref{ssec:corseis} does suggest that a background \PL\ \PS\ will make
a significant difference to the number of wave detections found in the
data.

All the automated detection algorithms described above attempt to find
individual pixels that are obviously oscillating at some frequency,
using the strength of the oscillation at that frequency (compared to
some assumption of the background) as the key quantity to measure.
These pixels are then clumped together and the quality of the signal
is then assessed.  The {\it wave coherence} detection approach of
\cite{2008SoPh..252..321M} implement wave detection through finding
spatially contiguous regions of the corona which are coherent in a
given frequency range.  {\BF \cite{2008SoPh..252..321M} use a
  symmetric filter to estimate the coherence, in that oscillatory
  power from either side of the central search frequency contributes
  the same weight to the final result.  The effect of this filter in
  the presence of a \PL\ \PS\ may be to bias detections to lower
  frequencies.}  Oscillations that are detected propagating along
loops are often found by considering distance-time plots
\citep{2000AA...355L..23D, 2003AA...404L...1K} and looking for
alternating brightening/darkening in those plots as a indicator of a
propagating wave.  This is essentially a coherence-based approach that
involves an implicit assessment of the likelihood that an oscillation
has been detected.  The basis of the assessment is a judgement that a
signal is observed in multiple contiguous pixels.  However, the results of
Section \ref{sec:anal} suggest that each pixel in these plots has a
background \PL\ \PS: the discussion of Section \ref{ssec:corseis}
therefore suggests that the significance of any claimed oscillation is
much reduced.

\subsection{``Excess'' emission}\label{ssec:excess}
The models fits of Section \ref{sec:anal} indicate that Model 2 is
preferred over Model 1 as the superior description of the observed
Fourier \PS. The moss region shows the greatest contribution from the
lognormal distribution $G(\nu)$ over the background \PL.
\cite{2005ApJ...624L..61D} show that it is possible to leak
oscillatory power from lower in the atmosphere to the upper reaches of
the atmosphere through the simple expedient of tilting the field line.
This changes the acoustic cutoff frequency and so allows wave energy
to propagate up in to the upper atmosphere.  If spectral model $M_{2}$
represents conditions in the solar atmosphere, then it should be
possible to infer a distribution of field line angles with the solar
surface using \cite{2005ApJ...624L..61D} and the observed power
distribution $G(\nu)$.  This would constrain the distribution of field
line angles in the lower atmosphere in different regions of the solar
surface.

Another possible cause of the excess emission is the appearance of
low-temperature emission lines in the AIA passbands.  Such lines are
more likely to be due to processes lower down in the solar atmosphere
where the amplitude of three and five minute oscillations is larger.
This is supported by three dimensional magnetohydrodynamic models of
the quiet Sun studied by \cite{2011ApJ...743...23M}.  It is shown that
lower temperature lines can be formed in quiet Sun areas that appear
in the AIA 171\AA\ and 193\AA\ wavebands. \cite{2011ApJ...743...23M}
find that the AIA 193\AA\ has more significant contributions from
non-dominant lines when compared to AIA 171\AA.  If this were the
case, then one would expect that the maximum of the ratio between
$G(\nu)/P_{1}(\nu)$ (see Figure \ref{fig:fit171193} and Table
\ref{tab:parameters}) would be higher for the AIA 193\AA\ \PS\ than
for the AIA 171\AA\ \PS, which is not observed.  However,
\cite{2011ApJ...743...23M} also note that their results are also quite
sensitive to the abundances used in their simulations, which may
explain the differences between their simulations and the observations
presented here.

\subsection{Nature of the \protect\PL\ \protect\PS}\label{ssec:nplps}
{\BF Power-law \PS\ have been
  observed in other solar phenomena.  As was mentioned above,
  \cite{2014AA...563A...8A} observe \PLs\ in the integrated
  emission of small portions of active regions and the quiet Sun as
  observed in the 195\AA\ passband images from EIT over the frequency
  range 0.01 - 1 mHz. Also, \citet{gupta2014} showed \PL\ \PA\ in the
  intensity at six single points in AIA 171\AA\ coronal plumes extending
  over the frequency range $0.3\rightarrow 4.0$ mHz.  Lower in the
  solar atmosphere, \cite{2008ApJ...683L.207R} show the presence of
  \PL\ Fourier \PA, in the range 7-20 mHz, in the Doppler velocity of
  the chromospheric Ca II 854.2 nm line as observed by the
  Interferometric Bidimensional Spectrometer (IBIS).  Further out in
  the solar atmosphere at 2.1 $R_{sun}$, \cite{2008ApJ...677L.137B}
  show the presence of \PL\ \PA\ in Ultraviolet Coronagraph
  Spectrometer observations of the intensity of Lyman-$\alpha$ in the
  frequency range $2.6\times10^{-6}\rightarrow1.3\times10^{-4}$ Hz.}

{\BF The coronal regions studied here also exhibit \PL-like \PA\ in the
approximate frequency range $0.05\rightarrow 10$ mHz.  Under the
modeling assumptions of models $M_{1}$ and $M_{2}$, the first
contribution to the observed power spectrum is a \PL\ (which is
detectable in all regions up to about 2 mHz)}.  It is present in both
wavebands studied and in all the regions, and so parsimony suggests
that the mechanism of its creation is both similar and ubiquitous in
all {\BF similar} parts of the solar atmosphere {\BF , that is quiet
  Sun, moss, loop footpoints and above sunspots}.  Further,
\cite{2014AA...563A...8A} show that a \PL\ \PS\ is present in
time-series of coronal active region and quiet Sun emission over the
frequency range 0.01 - 1 mHz, overlapping with the frequency studied
in this paper.  We hypothesize that the \PL\ \PS\ in the coronal
171\AA\ and 193\AA\ emission is due to the sum of a distribution of a
large number of events having different amounts of emission.
\cite{2011soca.book.....A} describes how a \PL\ \PS\ may obtained from
such a distribution, and that argument is outlined below.

Each emission event is modeled as an exponentially decaying function
of time $t$
\begin{equation}
\label{eqn:expdecay}
f(t) = \frac{E}{T}\exp\left(-\frac{t}{T}\right),
\end{equation}
for some timescale $T$ and emission $E$.  The corresponding \Fps\ is
\begin{equation}
\label{eqn:ftexpdecay}
P(\nu) = \frac{E}{1 + (2\pi \nu T)^{2}}.
\end{equation}
The total \Fps\ along the line of sight is given by
\begin{equation}
\label{eqn:sumftexpdecay}
P_{total}(\nu) = \sum_{T}N(T)P_{T}(\nu)
\end{equation}
where $N(T)$ is the distribution of time-scales for all the events
along the line of sight.  Further, if the number of events of a given
emission $E$ is assumed to be
\begin{equation}
\label{eqn:energydistrib}
N(E) \propto E^{-\alpha_{E}}
\end{equation}
and the total emission in each event depends on its time scale $T$
such that
\begin{equation}
\label{eqn:energytime}
E \propto T^{1+\gamma}
\end{equation}
then it can be shown that the observed \PS\ can be
approximated by
\begin{equation}
\label{eqn:finalfps}
P_{total}(\nu) \propto \nu^{-(2-\alpha_{E})(1+\gamma)}.
\end{equation}
This derivation shows it is possible to generate \PL\ power spectra
using swarms of statistically similar emission events.  {\BF Equation
  \ref{eqn:finalfps} extends a suggestion by
  \citet{1991SoPh..133..357H} that Equation \ref{eqn:ftexpdecay} be
  used to estimate the timescale of a number of smaller heating events
  with identical timescale $T$ in a sufficiently long span of solar
  X-ray observations. \citet{1991SoPh..133..357H} is extended by
  considering a distribution of heating event timescales (Equation
  \ref{eqn:energydistrib}) as opposed to a single timescale.}

The derivation of the \PL\ \PS\ Equation \ref{eqn:finalfps} shares
many similar features with the nanoflare model of {\it energy}
deposition in the solar atmosphere. In the nanoflare model advanced by
\cite{1988ApJ...330..474P}, large numbers of small magnetic
reconnection events convert the energy in the magnetic field into
energy that heats the corona.  Nanoflares are hypothesized to be small
($10^{24}-10^{27}$ ergs) yet ubiquitous in the corona.
\cite{2013ApJ...771...21W} and \cite{2013ApJ...770L...1T} analyzing
narrowband 193\AA\ images taken by the Hi-C instrument on board a
sounding rocket demonstrate the heating of small scale coronal
structures that appear to be consistent with scenarios of nanoflare
heating due to reconnecting magnetic loops.  However, the nanoflare
occurence rate, or their energy distribution for the entire solar
atmosphere is as yet unknown.

Any modeling or theoretical effort to infer the occurence rate and
energy distribution of energy deposition events from the observed
emission \PL\ \PS\ depends strongly on three factors that influence
the observation.  Firstly, many different loops lie along the
line-of-sight in the corona.  Each of these may have different
emission measures, temperatures and densities.  The location of energy
deposition may not be the same for each loop
\citep{2003A&ARv..12....1W, 2006SoPh..234...41K}.  This implies that
attempts to work back from the observed emission to the underlying
energy deposition must properly account for these line-of-sight
effects.

Secondly, the amount of energy and the wavelengths at which it is
radiated is a highly nonlinear function of the energy deposited in
each event, depending on the existing plasma properties and the
location of the energy deposition in the loop structure.  This must
all be consistently modeled on each loop and summed over multiple
loops to mimic the line-of-sight effects, as noted above.  For
example, \cite{2011ApJ...743...23M} use a full three-dimensional MHD
model to simulate an evolving solar atmosphere and its emission
\citep{2011AA...531A.154G}.  \cite{2008ApJ...682.1351K,
  0004-637X-752-2-161} have developed a zero-dimensional model of
individual strands in a coronal loop that describes the strand's
average temperature, pressure, and density.  Such models can be run
quickly and so can be used to model hundreds of coronal loops.

Finally, the AIA instrument filter response at the time of the
observation must also be understood and modeled {\BF in} order to
understand which emission lines contribute to the final observed
image, and ultimately the \Fps\ for each region.
\cite{2010AA...521A..21O} assume differential emission measure curves
for different regions on the Sun (active region, quiet Sun, coronal
hole, flaring flasma) and calculate synthetic spectra.  These
synthetic spectra were convolved with the effective area of each
channel, in order to determine the dominant contribution in different
regions of the solar atmosphere.  However, as noted by
\cite{2011ApJ...743...23M}, the differential emission measure curves
selected by \cite{2010AA...521A..21O} are not guaranteed to
accurately describe the conditions actually present, and so the
fractional contributions of each line in the AIA waveband are
estimates only.

These three factors will all have an effect on the inferred properties
of the energy deposition rate.  The observed \PL\ \PS\ here, coupled
with the emission event hypothesis and a modeling effort that takes
account of the factors above, offers a new way to constrain the energy
deposition rate in the solar corona.  Under the hypothesis, the energy
deposition rate can be explored through the distribution of the number
of events along the line-of-sight and their temporal dependence,
expressed through the parameters $\alpha_{E}$ and $\gamma$ in Equation
\ref{eqn:energydistrib}.

\section{Conclusions}\label{sec:conc}
Time series of AIA 171\AA\ and 193\AA\ images show approximately power
law like properties {\BF in four representative regions} of the
corona.  {\BF It has been shown that the detection of narrow band
  oscillatory power in the practice of coronal seismology must take in
  to account the presence of the \PL\ like \PS.  Detection thresholds
  should also increase in order to reduce the number of false positive
  detections in the presence of \PL\ like \PS.} The power spectra have
been modelled using three components: a \PL\ \PS, a lognormal
component, and a constant background.  It is hypothesized that the
\PL\ component is due to the sum along the line of sight of a
large number of statistically similar energy deposition events.  It is
also hypothesized that the lognormal component is due a combination of
the channeling of oscillatory power from lower to higher levels of the
solar atmosphere and lower temperature emission lines in the AIA
171\AA\ and 193\AA\ bandpasses.

Future studies will include examining power spectra at many more
locations on and off the disk. {\BF Automated definitions of coronal
  structures will be used to segment particular coronal features from
  the data, and so provide improved information of the differences the
  coronal structure makes to the observed \PL\ \PS.  Other AIA
  wavebands, such as AIA 355\AA, 94\AA, 131\AA and 211\AA\ will be
  used to look for further evidence of \PL\ \PA\ in different broad
  temperature ranges in the solar corona.}  The \PL\ index in
individual pixels will also be determined, and correlated with other
physical quantities such as the emission from those pixels.  Finally,
a better understanding of the source of the \PL\ \PS\
will come from comparing the \Fps\ of simulated observations to the
\Fps\ of the data in all the AIA wave-bands that are sensitive to
higher temperatures.  This will bring new constraints on the
simulations and improve our understanding of the temporal behavior of
the solar corona.



\acknowledgments

We are grateful to the developers of SSWIDL
\citep{1998SoPh..182..497F}, IPython \citep{ipython}, SunPy
\citep{mumford-proc-scipy-2013}, PyMC
\citep{Patil:Huard:Fonnesbeck:2010:JSSOBK:v35i04}, matplotlib
\citep{Hunter:2007} and the Scientific Python stack for providing data
preparation, manipulation, analysis and display packages.  This work
was supported by NASA award NNX13AE03G S01 funded through NASA ROSES
NNH12ZDA001N-SHP, and by a NSF Career grant 1255024 (JMA).



{\it Facilities:} \facility{SDO (AIA)}.



\appendix\section{Fitting models to the mean Fourier power spectra}\label{sec:app:ind}
In order to perform the fit, the noise in the \mFps\ (black lines,
Figure \ref{fig:fit171193}) must be estimated.  To do that, the
distribution of Fourier power in a region at each frequency must be
considered (see Figure \ref{fig:compare171193} for some example
distributions).  These distributions are constructed by calculating
the Fourier power at each pixel in the region, and binning those
powers by frequency.  However, the Fourier power at a given pixel is
not independent of the Fourier power at another pixel.  This is
because neighbouring pixels are highly likely to be physically
connected to each other, and so time-series in neighboring pixels are
likely to be strongly correlated (assuming that the background
structure exists over at least two pixels).  Also, the point spread
function of AIA spreads emission in one pixel over on to its
neighbors; therefore even neighboring pixels that are not physically
connected are still correlated.

Let the standard error in the mean of samples from a distribution be
given by $\sigma_{mean}$.  It is defined as the standard deviation of
the sampling distribution of the mean:
\begin{equation}
\sigma_{mean} = \sigma_{original} / \sqrt{N}
\label{eqn:sigmamean}
\end{equation}
where $\sigma_{original}$ is the standard deviation of the
distribution and $N$ is the sample size.  This formula assumes that
the measurements taken are all independent of each other, which is not
the case for the Fourier power spectra studied here.  Hence the value
of $N$ cannot be set to the number of pixels in a region.

Hence, the number of effectively independent pixels time-series must
be calculated. {\BF The dependence of pixels on each other is
  calculated using the following modeling assumptions.}  First, a pair
of randomly chosen next-nearest neighbor pixels are selected from the
region.  Next nearest neighbor pixels are chosen as close physical
proximity is most likely to express the strongest interdependence.
Next, the cross correlation $c(\tau)$ function between two randomly
chosen next-nearest neighbor pixels is calculated.  The cross
correlation coefficient measures the linear dependence of two time
series $X(t)$ and $Y(t \pm \tau)$ at different lag {\BF times $\tau$.
  The assumption here is that the linear dependence as measured by the
  cross-correlation coefficient estimates the actual dependence
  between time-series.}  The cross correlation coefficient is used to
define a coefficient of independence between the neighbors,
\begin{equation}
\label{eqn:ind}
\rho = 1 - \max|c(\tau)|.
\end{equation}
When $\rho$ is 1, the cross-correlation coefficient is zero, and there
is no measurable linear dependence between neighboring time series.
When $\rho$ is zero, the cross correlation coefficient is $\pm 1$, and
the two time-series are linearly dependent on each other.  This
procedure is repeated 10,000 times.  The effective number of pixels
$N_{eff}$ is defined as values is defined as
\begin{equation}\label{eqn:neff}
\label{eqn:nind}
N_{eff}= 1 + \hat{\rho}(N_{pixel}-1).
\end{equation}
where $\hat{\rho}$ is the arithmetic mean of the values of $\rho$
found.

\subsection{Model fitting and selection}
Equations \ref{eqn:pwrlaw} and \ref{eqn:pwrlawbump} are fit to the
geometric mean power spectra shown in Figures \ref{fig:fit171193}.  A
Bayesian/Markov chain Monte Carlo package (PyMC version 2.3,
\citealp*{Patil:Huard:Fonnesbeck:2010:JSSOBK:v35i04}) is used to construct
the Bayesian posterior and determine model parameter estimates and
their uncertainty.  If $P_{i,j}$ is the power at Fourier frequency
$\nu_{j}$ from Model $i$ then the likelihood $L$ is
\begin{equation}\label{eqn:likelihood}
L = \prod_{j=1}^{N_{f}}
\frac{1}{\sqrt{2\pi\sigma_{j}^{2}}}
\exp\left(
-\frac{(P_{i,j} - D_{j})^{2}}{2\sigma_{j}^{2}}
\right)
\end{equation}
where $D_{j}$ is the observed \mFps\ at frequency $\nu_{j}$, $N_{f}$
is the number of Fourier frequencies, and $\sigma_{j}$ is the normal
distribution width at frequency $\nu_{j}$.  {\BF Each of the priors
  for the parameters in each model has a uniform distribution, and are
  given in Table \ref{tab:priors}.  For some parameters, the logarithm
  of the model parameter has a uniform distribution, which aids in the
  exploration of parameter space}.  Each MCMC chain that was run took
100,000 samples, with burn-in assumed to be at 50,000 samples.  The
chain is thinned by taking every fifth element in the chain, forming
the final sample from which the parameter estimates and their
uncertainties are calculated.
\begin{deluxetable}{ccc}
\tabletypesize{\scriptsize} 
\tablecolumns{4}
\tablewidth{0pt}
\tablecaption{Priors for each of the parameters used in Model 1
  (Equation \ref{eqn:pwrlaw}) and
  Model 2 (Equation \ref{eqn:pwrlawbump}) \label{tab:priors}}
\tablehead{
\colhead{parameter} &
\colhead{Model 1 and 2} &
\colhead{Model 2 only} \\
}
\startdata
$\log_{10} A$ & $-4.34 \rightarrow 2.00$  & {\it not applicable} \\
$n$ & $1\rightarrow 6$ & {\it not applicable}  \\
$\log_{10} C$ & $-8.68 \rightarrow 2.00$ & {\it not applicable} \\
$\log_{10} \alpha$  & {\it not applicable}  & $8.68 \rightarrow -1.00$ \\
$\beta$ & {\it not applicable} & $0.1 - 10$ (mHz) \\
$\log_{10}\delta$ & {\it not applicable} & $0.00 \rightarrow 3.00$ \\
\enddata
\end{deluxetable}
The \mFpa\ are assumed to
be normally distributed with a width given by $\sigma_{mean}$ as
calculated above.  A five step fitting process is used to fit Models 1
and 2 to the data, and is described below.

\begin{enumerate}
\item The posterior is formed by multiplying the likelihood (equation
  \ref{eqn:likelihood}) by the priors for $A$, $n$ and $C$ (Table
  \ref{tab:priors}).  The MCMC sampler is run to fit Model 1, assuming
  $\sigma_{j} = \sigma_{original, j}$ the width of the distribution of
  Fourier power at frequency $G(\nu_{j})$.  Since the widths
  $\sigma_{original, j}$ are large it allows the fitting algorithm to
  converge to an approximate fit solution.
\item The posterior is formed by multiplying the likelihood (equation
  \ref{eqn:likelihood}) by the priors for $A$, $n$, and $C$ (Table
  \ref{tab:priors}).  The MCMC sampler is run to fit Model 1 with
  $\sigma_{j} = \sigma_{original, j}/sqrt{N_{eff}}$, seeded with
  estimated parameter values from step 1.  This generates a better
  estimate to the parameters of the model fit.  The resulting
  parameter estimates are given by the mean values of marginal
  probability distribution functions, and are the final estimates for
  Model 1.  The reduced chi-squared measurement $\chi^{2}_{r}$ for
  model $M_{1}$ is calculated using the values of $\sigma_{j}$ above.
\item The residuals of the model 1 fit to the data are then used to
  estimate the model parameters of $G(\nu)$.  This is done by
  (least-squares) fitting $G(\nu)$ to the positive residuals in the
  frequency range 0.1 - 10 mHz.
\item The MCMC sampler is run to fit Model 2 assuming $\sigma_{j} =
  \sigma_{original, j}$, seeded with the mean parameter values of
  Model 1 from step 2 and the estimated $G(\nu)$ model parameters from
  step 3.
\item The posterior is formed by multiplying the likelihood (equation
  \ref{eqn:likelihood}) by all the priors for Model 2 (Table
  \ref{tab:priors}).  The MCMC sampler is run to fit Model 2 with
  $\sigma_{j} = \sigma_{original, j}/\sqrt{N_{eff}}$, seeded with
  estimated parameter values from step 4.  The resulting parameter
  estimates are given by the mean values of marginal probability
  distribution functions, and are the final estimates for Model 2.
  The reduced chi-squared measurement $\chi^{2}_{r}$ for model $M_{2}$
  is calculated using the values of $\sigma_{j}$ above.
\end{enumerate}
{\BF The preferred model is selected using the Akaike Information
  Criteria \citep{2007MNRAS.377L..74L}, as described in the main
  text.}  Conveniently, the AIC is calculated by the PyMC package used
to perform the MCMC sampling.  The 68\% credible intervals for the
{\BF parameters of the model preferred using the AIC} are quoted in
Table \ref{tab:parameters}.  The upper and lower limits to the
credible intervals are found by determining the values beyond which
16\% of the probability lies in each wing of the marginal
distributions.




\bibliography{references}
\end{document}